\documentclass[12pt]{article}
\pdfoutput=1
\usepackage{amsmath,amsfonts,amssymb,amsthm,amssymb,bbm,bm}
\usepackage{pdfsync}
\usepackage{mathabx}
\usepackage{graphicx,import}
\usepackage[latin1]{inputenc}
\usepackage{hyperref}
\usepackage{empheq}
\usepackage[all]{xy}
\usepackage{stmaryrd}
\usepackage{rotating}
\usepackage{color}  
\usepackage{slashed}
\usepackage{cancel}
\usepackage{array, makecell} %
\usepackage{tikz-cd}
\usepackage{hhline}
\usepackage{comment}
\makeatletter

\DeclareFontFamily{U}{matha}{\hyphenchar\font45}
\DeclareFontShape{U}{matha}{m}{n}{
      <5> <6> <7> <8> <9> <10> gen * matha
      <10.95> matha10 <12> <14.4> <17.28> <20.74> <24.88> matha12
      }{}
\DeclareSymbolFont{matha}{U}{matha}{m}{n}
\DeclareMathSymbol{\oright}       {2}{matha}{"69}

\usepackage[mathscr]{euscript}

\newcommand{\doublehat}[1]{%
\begingroup%
  \let\macc@kerna\z@%
  \let\macc@kernb\z@%
  \let\macc@nucleus\@empty%
  \hat{\raisebox{.55ex}{\vphantom{\ensuremath{#1}}}\smash{\hat{#1}}}%
\endgroup%
}

\newcommand{\p}{\partial}

\newcommand{\bit}{\begin{itemize}}
\newcommand{\eit}{\end{itemize}}
\newcommand{\bd}{\begin{description}}
\newcommand{\ed}{\end{description}}

\newcommand{\bc}{\begin{center}}
\newcommand{\ec}{\end{center}}

\newcommand{\sG}{\mathfrak{g}}
\newcommand{\Tr}{\mathrm{Tr}}

\newcommand{\va}{\scriptscriptstyle}
\newcommand{\YM}{ {\cal R}}
\newcommand{\ym}{ r}
\newcommand{\YR}{  R}

\newcommand{\C}{{\mathbb C}}
\newcommand{\N}{{\mathbb N}}

\newcommand{\Z}{{\mathbb Z}}

\newcommand{\cQ}{{\mathcal{Q}}}

\def\be#1\ee{\begin{align}#1\end{align}}
\newcommand{\bea}{\begin{eqnarray}}
\newcommand{\eea}{\end{eqnarray}}
\newcommand{\bs}{\begin{subequations}}
\newcommand{\es}{\end{subequations}}

\newcommand{\la}{\label}

\newcommand{\tr}{{\rm Tr}}
\newcommand{\f}{\frac}

\newcommand{\bz}{{\bar{z}}}

\newcommand{\Ad}{\mathsf{Ad}}

\def\p{\partial}

\newcommand{\hC}{\widehat{C}}
\newcommand{\tN}{\widetilde{N}}

\def\d{\delta}

\def\rd{\mathrm{d}}
\def\pa{\partial }

\def\k{{\kappa^2} }

\newcommand{\tC}{{\tilde C}}

\def\hD{\hat{\Delta} }

\newcommand{\scri}{\cal I}

\begin{document}

\begin{titlepage}

\title{\Large{\bf On infinite symmetry algebras in Yang-Mills theory}}
\date{}
\maketitle
\thispagestyle{empty}

\vspace{-1cm}
\begin{center}
\author{\large{ Laurent Freidel$^1$,
Daniele Pranzetti$^{1,2}$,
Ana-Maria Raclariu$^{1,3}$
}
}

\vspace{1cm}

{\it $^1$Perimeter Institute for Theoretical Physics, 31 Caroline Street North, Waterloo, Ontario, Canada N2L 2Y5\\ \smallskip
\it $^2$ Universit\`a degli Studi di Udine, via Palladio 8,  I-33100 Udine, Italy\\ \smallskip
\it $^3$ Universiteit van Amsterdam, Science Park 904, 1098 XH Amsterdam, the Netherlands}
\end{center}

\vspace{0.5cm}

\begin{abstract}
Similar to gravity, an infinite tower of symmetries generated by higher-spin charges has been identified in  Yang--Mills theory by studying collinear limits or celestial operator products of gluons. This work aims to recover this loop symmetry in terms of charge aspects constructed on the gluonic Fock space. We propose an explicit construction for these higher spin charge aspects as operators which are polynomial  in  the gluonic annihilation and creation operators. The core of the paper consists of a proof that the charges we propose form a closed loop algebra to quadratic order. 
This closure involves using the commutator of the  cubic order expansion of the charges with the linear (soft) charge. Quite remarkably, this shows that this infinite-dimensional symmetry  constrains the non-linear structure of Yang--Mills theory. We provide a similar all spin proof in gravity for the so-called global quadratic (hard) charges which form the loop wedge subalgebra of $w_{1+\infty}$. 
\end{abstract}

\end{titlepage}

\tableofcontents

\section{Introduction}

It was shown in \cite{Guevara:2021abz} that gravity and gauge theories possess an infinite tower of symmetries generated by increasingly subleading soft modes. For example, the negative helicity modes organize in finite dimensional representations of the SL$(2,\mathbb{R})_L$ component of the Lorentz algebra (after analytic continuation to $(2,2)$ signature). In the case of pure Yang--Mills (YM) theory, the symmetry generators were related to the soft modes by a light-transform and were found to obey a simple algebra \cite{Strominger:2021mtt}\footnote{Note that this is the negative helicity version of the algebra worked out in \cite{Guevara:2021abz,Strominger:2021mtt}.}
\be 
\label{eq:current-algebra}
[S_m^{p,a}(\bz), S_n^{q,b}(\bz)] = -i f^{ab}_{~~c} S_{m+n}^{p + q - 1, c}(\bz).
\ee 
Here $p, q$ are half integers bigger than $1$ and $m,n$ satisfy the restriction 
$1-p\leq m\leq p-1$, while $m+p$ is restricted to be integer---and similarly for the pair $(q,n)$.
These soft modes admit a further mode expansion\footnote{ $p$ as defined in \cite{Strominger:2021mtt} is related to $s$ here via $p = \frac{s}{2} + 1$.}  \cite{Himwich:2021dau}
\be \label{modes}
S_m^{p,a}(\bz) =\sum_{n\in \mathbb{Z}} \bz^{\frac{s}{2} - n - 1} S_{m,n}^{p,a}.
\ee

The dictionary relating bulk asymptotic scattering states and operators in the celestial CFT \cite{Pasterski:2016qvg,Pasterski:2017kqt,Donnay:2020guq} suggests that celestial symmetries like \eqref{eq:current-algebra} should also be realized on the phase space of the theory. Progress in this direction was made in \cite{Freidel:2021ytz} where such a claim was established in Einstein gravity. Specifically, it was shown \cite{Freidel:2021qpz} that the asymptotic Einstein equations truncate to two towers of recursive differential equations for charges defined as appropriate combinations of the asymptotic metric components that transform covariantly under the homogeneous subgroup of the Weyl extension of BMS$_4$ \cite{Freidel:2021fxf}. After appropriate regularization, half of the non-linear charges (whose linear part corresponds to the same helicity gravitons) were shown to obey the gravitational analog of \eqref{eq:current-algebra}, namely a $w_{1 + \infty}$ algebra to linear order. Surprisingly, no restriction to the global subalgebra was necessary. One may therefore hope that the loop algebra continues to hold upon including higher non-linear contributions to the charges.

This paper  extends the analysis of \cite{Freidel:2021ytz} to Yang--Mills theory, beyond the linear level. In particular, we construct for each helicity an infinite tower of charges $R_s(\tau_s)$ labelled by a Lie algebra valued function $\tau_s(z,\bz) \in \sG $ on the celestial sphere $S$ (alternatively on the celestial torus $T=S^1\times S^1$). Here $(z,\bz)$ are complex  coordinates on $S$ (or circle coordinates on $T$). It can be shown that  $\tau_s$ is an element of an SL$(2,\mathbb{C})$ (or $\mathrm{SL}(2,\mathbb{R}) \times \mathrm{SL}(2,\mathbb{R})$) representation  of weight and spin $(\Delta,J) =(0,-s)$. These charges are constructed explicitly as operators acting on the Yang--Mills  Fock space and are obtained after regularization from components of the gauge field in a large-$r$ expansion. The latter satisfy a tower of recursive differential equations obtained from a large-$r$ expansion of the Yang--Mills equations
\be
\label{YM-diff-eq-intro}
\p_u \YM_{s} =  D \YM_{s - 1}+ i[A_z^{(0)}, \YM_{s - 1}]_{\sG} , \quad s \geq 0.
\ee
We prove that the charges $R_s(\tau_s)$ satisfy  the algebra 
\be \la{YM-alg-intro}
[R_s(\tau), R_{s'}(\tau')] = -g_{\va \rm{YM}}^2 R_{s+s'}([\tau,\tau']_{\sG}),
\ee 
up to quadratic order in the creation and anihilation operators and for \emph{arbitrary} functions  $\tau$ on $S$ (or $T$). We label the Lie algebra bracket with a subscript  $\sG$ to avoid any confusion with the quantum commutators of operators.

The global S-subalgebra \eqref{eq:current-algebra} appearing in celestial holography consists of the subalgebra generated by the $\tau_s$ solution of $D^{s+1}\tau_s =0 $, where $D$ is the covariant derivative with respect to $z$ on $S$ or $T$. The fact that this forms a subalgebra follows directly from the fact that $D^{s+s'+1}[\tau_s,\tau_{s'}]_{\sG}=0$, when $D^{s+1}\tau_s =0 $ and $D^{s'+1}\tau_{s'} =0 $. 
The algebra \eqref{eq:current-algebra} and its modes \eqref{modes} are recovered by choosing the smearing function on $T$ to be a polynomial of degree $s$ in $z$. More precisely,  for $\tau_s = {z^{m + \frac{s}{2}} \bz^{n - \frac{s}{2}} T^a} $, we have 
\be \la{eq:torus-modes}
S_{m,n}^{1+\frac{s}{2}, a} = \int_{T} \rd z \rd\bz    z^{m+ \frac{s}{2}} \bar{z}^{n - \frac{s}{2}}  r^a_s(z,\bz),
\ee
 where $r_s^a$ is the local charge aspect of spin $s$ and $T^a$ is a basis element for the Lie algebra $\sG$. The main result of the present work is that \eqref{YM-alg-intro} continues to hold for the \textit{local}, \textit{nonlinear} charges parameterized by arbitrary functions $\tau_s(z, \bz)$. In particular, at quadratic order, the commutator \eqref{YM-alg-intro} receives contributions from the \textit{cubic} component of the charges, 
\be 
\label{qcomm}
\begin{split}
[R_{s+}(\tau), R_{s'+}(\tau')]^2 &= [R_{s+}^2(\tau), R_{s'+}^2(\tau')] + [R^1_{s+}(\tau), R_{s'+}^3(\tau')] + [R_{s+}^3(\tau), R_{s'+}^1(\tau')],
\end{split}
\ee
where $R_s^k$ is the degree $k$ term in the charge regarded as a polynomial in the gauge fields, 
\be 
R_s(\tau) = \sum_{k = 1}^{s+2} R_s^k(\tau).
\ee
The last two terms on the right-hand-side of \eqref{qcomm} vanish for global $\tau_s$ but conspire to ensure that the quadratic order algebra \eqref{YM-alg-intro} is satisfied for arbitrary $\tau_s$. 

We also generalize the computation of the $w_{1 + \infty}$ phase-space algebra in \cite{Freidel:2021ytz} to the global quadratic charges in gravity. Finally, we explicitly show that the local spin-$2$ charges also obey a $w_{1 + \infty}$ algebra at the quadratic order. As in the YM case the inclusion of the cubic components in the charges is necessary to recover the correct commutation relations.

The relevance of the cubic component of the symmetry charge for the quadratic order $w_{1 + \infty}$ algebra for spin-2 charges in the matter sector was already pointed out in \cite{Hu:2022txx}. We are also aware of a forthcoming paper \cite{Hu:tocome} which presents complementary results on higher-helicity fields.

\section{Preliminaries}

We consider non-Abelian gauge theory with gauge group $G$ in 4-dimensional Minkowski spacetime. The Yang--Mills equations take the form
\be 
\label{eq:YM}
\rd \star F = 0,\quad F \equiv \rd {A} +i {A} \wedge {A},
\ee
where ${A}=A_\mu\rd x^\mu$ is a one-form valued in the adjoint representation of the Lie algebra $\mathfrak{g}$ of $G$. 

We begin by describing the construction of the charge aspects in terms of the asymptotic phase space variables of Yang--Mills theory. 
We work in Bondi coordinates\footnote{It can be convenient to work in retarded flat coordinates where $ds^2 = -2 du dr +  2r^2  d z d\bz$ and asymptotic infinity has the topology of a celestial plane. Removing the origin we get a celestial cylinder which can be compactified into a celestial torus $T$ with respect to which we express the mode expansions \eqref{eq:torus-modes}. } where 
\be 
ds^2 = -du^2 -2 du dr +  r^2  \frac{4 d z d\bz}{(1+|z|^2)^2},
\ee 
and assume an expansion of the field strength given by  
\begin{equation}
F_{ur} = \frac{1}{r^2}\sum_{n = 0}^{\infty} \frac{F_{ur}^{(n)}}{r^n}, \quad
F_{uz} = \sum_{n = 0}^{\infty} \frac{F_{uz}^{(n)}}{r^n}, \quad
F_{rz} = \frac{1}{r^2}\sum_{n = 0}^{\infty} \frac{F_{rz}^{(n)}}{r^n}, \quad
F_{z\bz} = \sum_{n= 0}^{\infty} \frac{F_{z\bz}^{(n)}}{r^n}. \label{radialexpF}
\end{equation}
In the radial gauge $A_r=0$, this corresponds to the following fall-off conditions on the components of the gauge potential
\begin{equation}
A_u=\sum_{n=0}^\infty\frac{A_u^{(n)}}{r^n},\qquad A_z=\sum_{n=0}^\infty\frac{A_{z}^{(n)}}{r^n}. \label{ansatzAs}
\end{equation}
We further specify the gauge on the initial slice to be such that $A_u^{(0)}=0$.

The radiative field which carries information about the gluonic creation and annihilation operators is given\footnote{In order to lighten the notation, we shall commonly indicate only $z$ in the functional dependence of the fields on the  coordinates on the sphere, e.g. $A_z^{(n)}(u,z)$, but it should be understood that in general the dependence is on both $z,\bz$.} by 
$ A_z^{(0)}(u,z)$.
 Our convention is such that the connection and field strength fields are hermitian. In other words we choose  $A_\mu =  A_\mu^a T_a$, 
where $T_a $ is a Hermitian generator satisfying the algebra $[T_a,T_b]=i f_{ab}{}^c T_c.$\footnote{Our conventions here are such that the structure constants differ by a sign compared to those in \cite{He:2015zea,Pate:2017vwa}.} The trace is normalized to $\Tr(T_aT_b)=\delta_{ab}.$

The first few charge aspects are identified as the dominant elements in the radial expansion of the field strength:
\be \la{recYMbc}
 \YM_{-1} = F_{\bz u}^{(0)}, \qquad\YM_{ 0} = \frac{1}{2}\left(F_{ru}^{(0)} +  F_{\bz z}^{(0)}\right), \qquad  \YM_{1}= F_{r z}^{(0)}.
\ee
The higher spin charge aspects $\YM_{s}$ of conformal dimension and spin $(\Delta,J)=(2,s)$ are constructed recursively by solving a system of  differential equations given by 
\be
\la{recYM}
\p_u \YM_{s} =  D \YM_{s - 1}+ i[A_z^{(0)}, \YM_{s - 1}]_{\sG} , \quad s \geq 0.
\ee 
These evolution equations are consequences of the Yang--Mills evolution for $\YM_0$ and $\YM_1$. For higher spin $s\geq1$ they correspond to a truncation of the full Yang--Mills equations expanded in $1/r$.
These equations  parallel the ones extracted from the asymptotic Einstein equations in \cite{Freidel:2021ytz, Freidel:2021qpz}.
A complete derivation of this result starting from the Yang--Mills equations \eqref{eq:YM} will be provided elsewhere \cite{NicoandI}.  The spin-$0$ charge is the leading while the spin-$1$ charge is the subleading one.

The recursion relations \eqref{recYM} are formally solved in terms of $\YM_{-1}$ by
\be 
\label{YM-charges}
\YM_s = (\p_u^{-1}[D+i\Ad(A_z^{(0)}) ] )^{s +1} \YM_{-1},
\ee
where $\Ad(X)Y=[X,Y]_{\sG}$ denotes the adjoint action and $(\pa_u^{-1} O)(u) :=\int_{+\infty}^u \rd u' O(u')$ for functionals that satisfy the boundary condition $O(+\infty)=0$.
This expression is  non-linear in the radiation field $A$ and it will therefore be convenient to expand the charge aspects as
\be
\YM_s(u,z)&=\sum_{k=1}^{s+2}\YM^k_{s}( u,z)\,,
\ee
where $\mathcal{R}^k_s$ is homogeneous of degree $k$\footnote{More precisely it is of degree $1$ in ${A^*}$ and degree $k-1$ in $A$} in the gauge fields $A, {A^*}$. In the following we use that $\mathcal{R}^k_s=0$ unless $k \leq s+2$.
At linear order we simply have
\be
\YM^1_s(u,z)&=(\p_u^{-1}{D})^{s+1 } F_{\bz u}^{(0)}(u,z)\,,
\ee
while the higher order components are recursively determined in terms of the lower order ones by 
\be\la{Rk}
\YM^k_{s}(u, z)= i\sum_{n={k-2}}^s  (\p_u^{-1})^{s-n+1} D^{s-n} \left[A_z^{(0)}(u,z), \YM^{k-1}_{n-1}(u,z)  \right]_{\sG}\qquad  {\rm for}\qquad k\geq 2\,.
\ee
It turns out that \eqref{Rk} suffer from divergences in the limit $u \rightarrow -\infty$, the past boundary $\mathcal{I}^+_-$ of future null infinity $\mathcal{I}^+$. This can be remedied by defining the renormalized charge aspects
\be\la{rk}
 \ym^k_s(z)&:= \lim_{u\to-\infty} \sum_{\ell=0}^{s} \frac{(-)^{s-\ell}u^{s-\ell}}{(s-\ell)!} D^{s-\ell} \YM^k_\ell(u,z),
\ee
whose action on the corner phase space at $\mathcal{I}^{+}_-$ is finite. The first two non-linear components of these charges will be central to our analysis and are given explicitly by (see Appendix \ref{App:YMcharges})
\be
 \ym^1_s(z) &= (-)^{s+1} \int_{-\infty}^\infty  \frac{u^s}{s!}  D^{s+1} F_{\bz u}^{(0)}(u,z)\,,\la{r1}\\
  \ym^2_s(z)&= -i  \sum_{n=0}^{s}  \int_{-\infty}^\infty 
 \frac{(-u)^{s-n}}{(s-n)!} D^{s-n} \left[A_z^{(0)}(u,z), (\p_u^{-1} D)^{n} F_{\bz u}^{(0)}(u,z) \right]_{\sG}\,,\la{r2}\\
  \ym^3_s(z)&=\sum_{n=1}^{s}  \sum_{k=0}^{n-1}\int_{-\infty}^\infty\rd u
 \frac{(-u)^{s-n}}{(s-n)!} \cr
 &\times
 D^{s-n}  \left[ A_z^{(0)}(u,z) , (\p_u^{-1})^{k+1} D^k  \left[ A_z^{(0)}(u,z), (\p_u^{-1}D)^{n-k-1}  F_{\bz u}^{(0)}(u,z)\right]_{\sG}\right]_{\sG}.\la{r3}
 \ee  

It will prove useful to express these charge aspects in a discrete basis where all the integrals over $\scri$ disappear, in analogy to the gravitational case recently considered in \cite{Freidel:2022skz}. We perform the discrete basis charge construction in Section \ref{sec:YM-charges} and use it in Section \ref{sec:YM-alg} to derive the algebra \eqref{YM-alg-intro}.

Motivated by the holographic calculation in \cite{Guevara:2021abz}, the linear component of the analogous commutator in gravity was first computed in  \cite{Freidel:2021ytz}. A simpler derivation will be given in Section \ref{sec:GR-alg}  using the expansion of the asymptotic fields in terms of a discrete tower of modes recently derived in \cite{Freidel:2022skz} (see also \cite{Cotler:2023qwh} for a complementary analysis) and reviewed in Sections \ref{sec:GR} and \ref{sec:GR-charges}, where the cubic charges are derived for the first time; in addition, we will also provide evidence for the gravitational $w_{1+\infty}$ loop algebra  at quadratic order as well.

\subsection{Discrete basis}\la{YM-db}

In order to introduce the discrete basis, we first decompose  the vector potential in terms  the positive and negative energy fields defined as 
 \be
A_+(u,z) := \frac1{2\pi} \int_0^\infty \rd \omega   e^{-i\omega u} \widetilde{A}_+(\omega,z)
=-\frac1{2i\pi}  \int_{-\infty}^{+\infty} \rd u' \frac{A_z^{(0)}(u',z)}{(u'-u+i\epsilon)}, \cr
 A_-(u,z) := \frac1{2\pi} \int_0^\infty \rd \omega   e^{-i\omega u}  \widetilde{A}_-(\omega,z)
=-\frac1{2i\pi}  \int_{-\infty}^{+\infty} \rd u' \frac{A_z^{(0)*}(u',z)}{(u'-u+i\epsilon)}.
\ee 
These fields enter the decomposition of the leading
components $A^{(0)}$ and $F^{(0)}$ as follows\footnote{Note that $F_\pm(u,z)=-\pa_u A_{\pm}(u,z)$.}
 \be
 A_z^{(0)}(u,z)&
{  = A_+(u,z)+A_-^*(u,z)}\,,\\
 F_{\bz u}^{(0)}(u,z)&
 { = F_-(u,z)+ F_+^*(u,z)}\,.
 \ee
 We introduce also the Mellin transforms $\widehat{A}_{\pm}(\Delta)$ and $\widehat{F}_{\pm}(\Delta)=i \widehat{A}_{\pm}(\Delta+1)$ of $\widetilde{A}_{\pm}(\omega)$ and  $\widetilde{F}_{\pm}(\omega)$ respectively
\be 
\label{Mellin}
\widehat{A}_\pm(\Delta) &:= \int_{0}^{+\infty} \rd\omega \omega^{\Delta-1} \widetilde{A}_\pm(\omega).
\ee
This equation implies that
\be 
\widehat{F}(\Delta) = -i^\Delta \Gamma(\Delta+1) \int_{-\infty}^{\infty} \rd u (u+i\epsilon)^{-(\Delta+1)} A(u) = i^\Delta \Gamma(\Delta) \int_{-\infty}^{\infty} \rd u (u+i\epsilon)^{-\Delta} F(u).
\ee
By demanding the vector potential field to  belong to the Schwartz space $\mathcal{S}$ \cite{Schwartz}, we can then introduce 
the YM {\it memory observables}
\be\la{Fn}
 \mathscr{F}_\pm(n) := \mathrm{Res}_{\Delta = -n} \widehat{F}_{\pm}(\Delta), \quad n \in \mathbb{Z}_+\,.
 \ee
These can be computed from the integrals
 \be \label{softF}
 \mathscr{F}_+(n) 
 : = \lim_{\omega \rightarrow 0^+} \frac{i^n}{n!}\left( \int_{-\infty}^{+\infty} \rd u\,e^{i\omega u} u^n F^{(0)}_{z u} (u)\right) = \frac{i^n}{n!} \left( {\oint}_U\rd u\, u^n  F_{+}  (u)\right),
 \ee
where in the first integral we take the limit  $\omega \rightarrow 0$ from above and in the second  $U$ is the  upper half plane contour.  The negative modes $\mathscr{F}_-(n)$ are defined by similar integrals but with $ F^{(0)}_{z u}$ replaced by  $F^{(0)}_{\bz u}$. These memory observables can be understood as the coefficients in a Taylor expansion  of $\widetilde{F}_\pm(\omega)$ around $\omega=0$, namely
 \be
 \label{FFourier}
 \widetilde{F}_{\pm}(\omega)=\sum_{n=0}^\infty  \mathscr{F}_\pm(n) \omega^n, \qquad 
 \mathscr{F}_\pm(n) = \frac{1 }{n!} \left.
 \pa^n_\omega  \widetilde{F}_{\pm}(\omega)\right|_{\omega=0^+} .
 \ee

At the same time, the Goldstone fields are defined by evaluating $\widehat{F}_{ \pm}(\Delta)$ at positive integer $\Delta$, namely
\be 
\label{An}
\mathscr{A}_{\pm}(n) :=\lim_{\Delta \rightarrow n}\widehat{F}_{\pm}(\Delta), \quad n \in \mathbb{Z}_+\,.
\ee
They correspond to Taylor coefficients in the analytic expansion of $A_{\pm}(u)$ around $u=0$
\be\label{AA}
\left.  \pa_u^n A_{\pm}(u)\right|_{u=0} =
\f{i^{-1- n}}{2\pi } \mathscr{A}_\pm(n)\,,\qquad
A_{\pm} (u)
=\f{1}{2i\pi }  \sum_{n=0}^\infty \frac{(-iu)^n}{n!}   \mathscr{A}_\pm(n)\,.
\ee
Following the gravity analysis in \cite{Freidel:2022skz}, it can be shown that \eqref{Fn} and \eqref{An} form a basis for asymptotic gauge potentials that belong to the Schwartz space.

\subsection{Phase space}

The YM phase space at asymptotic infinity is characterized by the symplectic potential
\be
\Theta^{\va \rm{YM}}=\f1{g_{\va \rm{YM}}^2}\int_{\scri^+} {\rm Tr}\left[F_{\bz u}^{(0)} (u,z) \d A_z^{(0)}(u,z)  +F_{z u}^{(0)} (u,z)\d A_{\bz}^{(0)}(u,z)\right]\,,
\ee
where $\Tr$ denotes the Cartan--Killing form for the Lie algebra associated to the YM theory.  
Modulo a canonical transformation, this can be rewritten as $\Theta^{\va \rm{YM}}=\Theta^{\va \rm{YM}}_++\Theta^{\va \rm{YM}}_-$ where
\be
\Theta^{\va \rm{YM}}_\pm=\f1{g_{\va \rm{YM}}^2}\int_{\scri^+}  {\rm Tr}\left [F_{ \pm} (u,z) \d A_{\pm}^{*}(u,z)\right]\,.
\ee
By means of \eqref{softF}, \eqref{AA}, we can rewrite the two symplectic potential components in terms of the YM memory and Goldstone modes as
\be
\Theta^{\va \rm{YM}}_\pm=\f1{2\pi i g_{\va \rm{YM}}^2}\sum_{n=0}^\infty \int_S {\rm Tr}\left[ \mathscr{F}_\pm(n,z) \d \mathscr{A}^*_{\pm}(n,z)\right].
\ee
In the quantum theory, the only non-trivial commutator is then given by
\be\la{com}
[ \mathscr{F}^a_\pm(m,z),  \mathscr{A}^{b\dagger}_{\pm}(n,z')]=2\pi g_{\va \rm{YM}}^2 \d^{ab}\delta_{n,m}\d^2(z,z').
\ee

\section{YM corner charges}\la{sec:YM-charges}

In this section we rewrite YM higher spin charge operators in terms of the soft variables introduced in the previous section. We then compute their action on the discrete modes and review the connection with the celestial OPE \cite{Pate:2019lpp,Guevara:2021abz}. Finally we demonstrate that the global subalgebra of quadratic charges is precisely \eqref{YM-alg-intro}.

\subsection{Charge aspects}

All the charge operators of level $k$ can be decomposed  as sum of a positive helicity charge and the conjugate of a negative helicity charge operator according to 
\be
\ym^k_s(z)=  \ym^{k}_{s+}(z)+ \ym^{k\dagger}_{s-}(z). 
\ee
The decomposition of the linear, quadratic and cubic charges follows straightforwardly from \eqref{r1},\eqref{r2},\eqref{r3}.
One finds that 
\be\la{r1n}
  \ym^1_{s\pm}(z) 
&=-i^{-s} D^{s+1} \mathscr{F}^\dagger_\pm(s,z)\,.
\ee
Similarly for the quadratic and cubic charges one finds that 
(see Appendix \ref{App:YMcharges-cor})
\be\la{r2n}
 \ym^2_{s\pm}(z)&=-\f{i^{-s}}{2\pi} \sum_{\ell=0}^\infty \sum_{n=0}^{s} 
  \left(\begin{matrix}
\ell+n \\
\ell
\end{matrix}\right)
 D^{n} \left[\mathscr{A}_\pm(\ell,z), D^{s-n}  \mathscr{F}^\dagger_\pm(s+\ell,z) \right]_{\sG},\\
\la{r3n}
 \ym^3_{s\pm}(z)
 &=
-
 \f{i^{-s}}{(2\pi)^2}
 \sum_{\ell=0}^\infty \sum_{m=0}^\infty  \sum_{n=0}^{s-1}  \sum_{k=n}^{s-1} 
 \left(\begin{matrix}
\ell+n \\
\ell
\end{matrix}\right)
 \left(\begin{matrix}
m+ \ell +k+1 \\
m
\end{matrix}\right)
 \cr
 &\times
  D^{n}  \left[ \mathscr{A}_\pm(\ell,z)  ,   D^{k-n}  \left[ \mathscr{A}_\pm  (m,z) ,
 D^{s-k-1} \mathscr{F}^\dagger_\pm(s+\ell+m,z)\right]_{\sG}\right]_{\sG}\,.
 \ee
 These charge aspects are valued in the Lie algebra $\sG$.

\subsection{Charges}\la{sec:YMcharges}

Given the charge aspects expressed as corner integrals of the memory and Goldstone variables, we introduce the symmetry charges labeled by Lie algebra valued generators of spin $-s$ denoted $\tau_s=\tau_s^a T_a$ \footnote{We define $\int_S:=\int_S \rd^2z \sqrt{\gamma}$, with $\gamma$ the determinant of the 2-sphere metric, and $\int_{\scri^+}:= \int_{-\infty}^\infty \rd u \int_S$. When different coordinate systems on the 2-sphere $z,z'$ are introduced, we make the measure explicit to avoid confusion.}
\be\la{Qs}
\YR_{s\pm}(\tau):= \int_{S}  \Tr\left(\tau(z) \ym_{s,\pm  }(z)\right)= \int_{S}  \tau_{a}(z) \ym_{s\pm  }^a(z)\,.
\ee
Explicitly, the positive/negative helicity linear, quadratic and cubic charges read
\be
\YR^{1}_{s\pm}(\tau)&=  i^{s}\int_{S}  { {\rm Tr}}\left(D^{s+1} \tau(z)    \mathscr{F}^{\dagger}_\pm(s,z)\right)\,,\la{YR1}\\
\YR^{2}_{s\pm}(\tau)&= \f{i^{-s}}{2\pi}   \sum_{\ell=0}^\infty \sum_{n=0}^{s} 
(-)^n
  \left(\begin{matrix}
\ell+n \\
\ell
\end{matrix}\right)
\int_{S}  { {\rm Tr}}\left(D^{n}\tau(z) 
\left[ D^{s-n}  \mathscr{F}^\dagger_\pm(s+\ell,z),\mathscr{A}_\pm(\ell,z)\right]_{\sG}\right),\la{R2-1}
\\
\YR^{3}_{s\pm}(\tau)&=
-
 \f{i^{-s}}{(2\pi)^2} \sum_{\ell=0}^\infty \sum_{m=0}^\infty  \sum_{n=0}^{s-1}  \sum_{k=n}^{s-1}
 (-)^{n}
 \left(\begin{matrix}
\ell+n \cr
\ell
\end{matrix}\right)
 \left(\begin{matrix}
m+\ell+k +1\\
m
\end{matrix}\right)
 \cr
 &\times
 \int_{S} { {\rm Tr}}\left( D^{n}  \tau(z)
\left[ \mathscr{A}_\pm(\ell,z)  ,   D^{k-n}  \left[ \mathscr{A}_\pm  (m,z) ,
 D^{s-k-1} \mathscr{F}^\dagger_\pm(s+\ell+m,z)\right]_{\sG}\right]_{\sG}\right)\,.\la{YR3}
\ee

\subsection{Quadratic charge action}
The action of the quadratic charge operator can be conveniently written in terms of the Lie algebra valued operator
\be\la{Ps}
P^a_s(\alpha; \tau):=  \sum_{n=0}^{s}  \f{(-\alpha)_n}{n!} (D^{n}\tau^a_s(z))  D^{s-n}.
\ee
An essential property of this operator, proven in Appendix \ref{App:YMcoch} concerns its behavior under transposition. Given two lie algebra functions $(A,B)$ on $S$, it satisfies 
\be 
\int_{S} {\rm Tr} \left( 
 [P_s(\ell+1; \tau),  B(z)]_{\sG} \,A(z) \right)=
 -(-1)^s 
 \int_{S}  
 {\rm Tr}\left(   
 \left[P_s(-(s+\ell+1); \tau), A(z)\right]_{\sG}   B(z) 
\right).
\label{Transpose}
\ee 
This implies that the quadratic charge can be conveniently written in terms of this operator as 
(see Appendix \ref{App:YMcoch})
\be\la{YR2c}
\YR^{2}_{s\pm}(\tau)&= 
  \f{i^{-s}}{2\pi} \sum_{\ell=0}^\infty  
\int_{S} {\rm Tr} \left( 
 [P_s(\ell+1; \tau),  \mathscr{F}^{\dagger }_\pm(s+\ell,z)]_{\sG} \,\mathscr{A}_\pm(\ell,z) \right),\cr
 &= -\f{i^{s}}{2\pi}   \sum_{\ell=s}^\infty 
\int_{S}  
 {\rm Tr}\left(   
 \left[P_s(-(\ell+1); \tau), \mathscr{A}_\pm(\ell-s,z)\right]_{\sG}   \mathscr{F}^\dagger_\pm(\ell,z) 
\right),
 \ee
where $(x)_n= x(x-1)\cdots(x-n+1)$ denotes the falling factorial.

From the expressions \eqref{R2-1} and \eqref{YR2c} for the charges we can evaluate the quantum commutator of charges when acting on the discrete fields. One finds that 
\be
[\YR^{2}_{s\pm}(\tau), \mathscr{F}^{\dagger b}_{\pm}(n,z)] &=i^{-s}g_{\va \rm{YM}}^2 \left[ 
 P_s(n+1; \tau), \mathscr{F}^{\dagger }_{\pm}(s+n,z)\right]_\sG^{b}\la{R2Fb}\\
 [\YR^{2}_{s\pm}(\tau), \mathscr{A}^{b}_\pm(n,z)] &= i^{ s} g_{\va \rm{YM}}^2  
 \left[P_s(-(n+1); \tau),\mathscr{A}_+(n-s,z)\right]_\sG^{b}.\la{R2Ab}
 \ee 
 The quantum commutators of $\YR^{2}_{s\pm}(\tau)$ with $\mathscr{A}^{b}_\mp$ and $\mathscr{F}^{\dagger b}_{\mp}$ obviously vanish.
 
 Given that $\mathscr{A}^{b}_+(n,z) \propto \lim_{\Delta= 1+n} \hat{A}^b(\Delta) $ and that $\mathscr{F}^{b}_+(n,z) \propto \mathrm{Res}_{\Delta= 1-n} \hat{A}^b(\Delta) $ one can deduce from this the action on the  Mellin transform of the asymptotic field. It is simply given by 
\be\label{YRC}
[\YR^{2}_{s\pm}(\tau), \hat{A}_\pm^{\dagger b}(\Delta)]&= i^{-s} g_{\va \rm{YM}}^2
\left[P_s(-\Delta+2; \tau), \hat{A}_\pm^\dagger(\Delta-s)\right]_\sG^{b},\\
[\YR^{2}_{s\pm}(\tau), \hat{A}_\pm^b(\Delta)]&=i^{ s} g_{\va \rm{YM}}^2
\left[P_s(-\Delta; \tau), \hat{A}_\pm(\Delta-s) \right]_\sG^{b}. \label{YRC2A}
\ee

\subsection{Celestial OPE from charge action}

From \eqref{YRC} we get that the commutator between the charge aspect and the Mellin transform of the radiative field is 
\be 
\left[ \ym_{s+}^{2a}(z_1),\hat{A}_+^{\dagger b}(\Delta_2,z_2) \right]
=i^{1-s} g_{YM}^2 f^{ab}{}_c 
\sum_{n=0}^s (-1)^{s-n}
\frac{(\Delta_2 -2)_{s-n}}{(s-n)!}
\pa_1^{s-n} \delta^{(2)}(z_{12}) \pa_2^n \hat{A}_+^{\dagger c}(\Delta_2-s,z_2) .
\ee
The correspondence between the Fock space commutator and the OPE is obtained through the identification 
\be 
\ym_s^{1a}(z_1) \hat{A}_+^{b\dagger}(\Delta_2,z_2) 
\sim 
-\frac{1}{2}\left[ \ym_{s+}^{2a}(z_1),\hat{A}_+^{b \dagger}(\Delta_2,z_2) \right].
\ee 
Now given that
\be  
\ym^1_{s\pm}(z) 
=-i^{-s} \pa_z^{s+1} \mathscr{F}^{\dagger}_\pm(s,z), \qquad 
\pa_z^{s+1} \frac1{n!}
\left( \frac{z^n}{\bar{z}}\right)=2\pi \pa_z^{s-n}\delta^{(2)}(z),
\ee 
we obtain the OPE 
\bea\mathscr{F}^{\dagger a}_{+}(s,z_1) 
\hat{A}_+^{\dagger b}(\Delta_2,z_2) \sim  
 i\frac{g_{YM}^2}{4\pi} \frac{f^{ab}{}_c }{\bz_{12}}
\sum_{n=0}^s {(-1)^{s-n}}
\frac{(\Delta_2 -2)_{s-n}}{(s-n)!} \frac{z_{12}^n\pa_2^n}{n!} \hat{A}_+^{\dagger c}(\Delta_2-s,z_2).
\eea 
 Finally, one uses that $\mathscr{F}^{\dagger a}_\pm(s,z) =  {-i}\mathrm{Res}_{\Delta= 1-s} \hat{A}_{\pm}^{\dagger a}(\Delta)$ and the evaluation 
$\mathrm{Res}_{\Delta= 1-s} \Gamma(\Delta -1 +n )= \frac{(-1)^{s-n}}{(s-n)!}$ for $n \leq s $, to see that the previous OPE is the residue at $\Delta_1=1-s$ of 
\be
\hat{A}_+^{\dagger a}(\Delta_1,z_1) 
\hat{A}_+^{\dagger b}(\Delta_2,z_2) \sim  
- \frac{g_{YM}^2}{4\pi} \frac{f^{ab}{}_c }{\bz_{12}}
\sum_{n=0}^s 
\frac{\Gamma(\Delta_1 -1 +n ) \Gamma(\Delta_2-1)}{\Gamma(\Delta_1 + \Delta_2 + n - 2) } \frac{z_{12}^n\pa_2^n }{n!}\hat{A}_+^{\dagger c}(\Delta_1+\Delta_2-1,z_2).
\ee 
This is the complex conjugate of the tree level OPE for positive-helicity gluons derived in \cite{Pate:2019lpp, Guevara:2021abz}.\footnote{Recall that the celestial operators $\mathcal{O}^+_{\Delta}$ are related to $\hat{A}^{\dagger a}_+$ via $\mathcal{O}^+_{\Delta} = \frac{4\pi i}{g_{YM}} \hat{A}^{\dagger a}_+$.}

\subsection{Global charge}

YM {\it global charges} are characterized by the condition
\be\la{YMGcon}
D^{s+1} \tau_s(z)=0.
\ee
By means of the generalized Leibniz rule\footnote{ This follows from the expansion $(x+y)^{\alpha} =  \sum_{n=0}^\infty \frac{ (\alpha)_{n}}{ n! }  x^n y^{\alpha-n} $ valid when  $x<y$.}
\be
 D^\alpha  \tau^a_s  D^{s-\alpha} 
 =\left[ \sum_{k=0}^s +\sum_{k=s+1}^\infty \right]  \frac{(\alpha)_{k}}{ k! }   (D^{k} \tau^a_s ) D^{s-k}\,,\la{magicYM}
\ee
we conclude that, for the global charges, the second sum in \eqref{magicYM} drops out and we can write the  operator \eqref{Ps} simply as a conjugation 
\be\la{Pglob}
P^a_s(-\alpha; \tau_s)= D^{\alpha} \tau^a_sD^{s-\alpha} \,.
\ee
where the product  is simply the composition of the operations of differentiation and multiplication by $\tau_s$. This means that we can write the quantum commutator \eqref{YRC2A} in terms of the adjoint action defined around \eqref{YM-charges} as
\be 
[\YR^{2}_{s\pm}(\tau), \hat{A}_\pm^b(\Delta)]&=i^{ s} g_{\va \rm{YM}}^2
\Ad[P_s(-\Delta; \tau)] \hat{A}_\pm(\Delta-s) 
= {i^{ s} g_{\va \rm{YM}}^2} D^{\Delta}\Ad[\tau] D^{s-\Delta}\hat{A}_\pm(\Delta-s) .
\ee
From this we see that the double quantum commutator action on $\hat A(\Delta)$ is given by
\be
[\YR^{2}_{s\pm}(\tau),[ \YR^{2}_{{s'}\pm}(\tau'),\hat{A}_\pm^b(\Delta)]]&=i^{ s+s'} g_{\va \rm{YM}}^4
\Ad[P_{s'}(-\Delta; \tau')]\Ad[P_{s}(s'-\Delta; \tau)]  \hat{A}_\pm(\Delta-s-s').
\ee 
When $\tau$ and $\tau'$ are parameters of global symmetry we simply have that 
\be
\Ad[P_{s'}(-\Delta; \tau' )] \Ad[P_{s}(s'-\Delta; \tau)]
&= D^{{\Delta}} \Ad[\tau'] \Ad[ \tau] D^{s+s'-{\Delta}}.
\ee
The antisymmetrization of this action  with respect to $s\leftrightarrow s'$ gives the action of the commutator $[\YR^{2}_{s\pm}(\tau), \YR^{2}_{{s'}\pm}(\tau')]$ on $\hat{A}(\Delta)$.
Using that $[\Ad[\tau'] ,\Ad[ \tau] ] = \Ad([\tau', \tau])$, we thus obtain that the algebra of \emph{global} charges satisfies for each helicity the global S-algebra:
\be 
[\YR^{2}_{s\pm}(\tau), \YR^{2}_{{s'}\pm}(\tau')] = -g_{\va \rm{YM}}^2 \YR^{2}_{s+s', \pm}([\tau, \tau']_\sG).\la{GlobalCharge}
\ee 
We can also conclude from our definition that the commutator of global charges of opposite helicities commute
\be 
[\YR^{2}_{s\pm}(\tau), \YR^{2}_{{s'}\mp}(\tau')]=0.
\ee 
Since the total charge is the sum $R_s(\tau)= R_{s+}(\tau) + R_{s-}^\dagger(\tau)$ we have that the charge algebra for 
$R_s$ is identical to \eqref{GlobalCharge}.

On the torus $T$, the global algebra is a loop algebra parametrized by global charge parameters $\tau (n,m)= z^{m + \frac{s}{2}} \bz^{n - \frac{s}{2}} T^a $, where $n,m \in \mathbb{N}$ and $-\frac{s}{2} \leq m \leq \frac{s}{2}$. In this case we define $S^{\tfrac{s}{2}+1,a}_{m,n}:=\YR^{2}_{s\pm}(\tau(n,m))$. This is the algebra revealed by \cite{Strominger:2021mtt} from the study of the OPE. It is the analog for Yang--Mills of the w$_{1+\infty}$ loop algebra. 
This global algebra also arises naturally in the study of self-dual Yang--Mills in the twistor formulation \cite{Adamo:2021zpw,Adamo:2022wjo}.

It is important to appreciate that on the complex sphere $S^2$ the set of global charges vanish if we insist that $\tau$ is a regular function on $S^2$.\footnote{The reason is that $\tau_s$ is a spin $-s$ function on the sphere and therefore it can be expanded in spin spherical harmonics $Y^{-s}_{\ell m}$ with $\ell\geq s$ \cite{goldberg1967spin}. We have that $D^{s+1}Y^{-s}_{\ell m}= \sqrt{\frac{(\ell+s)!}{(\ell-s)!}} DY^0_{\ell m}$, where $Y^0_{\ell m}= Y_{\ell m}$ are the usual scalar spherical harmonics. From this we see that the global condition reduces to $DY_{\ell m}=0$ which can only be satisfied for $\ell=0$ hence for $s=0$ since $s\leq \ell$.} Non-trivial charges can be obtained by allowing for poles in $\tau$ at isolated points on the sphere and \eqref{YMGcon} will only hold away from these points. As we will see, the non-linear contributions to the charges such as \eqref{YR3} will be crucial in this case to ensure that the charge algebra closes.


\section{YM corner algebra for the local charges}\la{sec:YM-alg}

Given the charges derived in Section \ref{sec:YMcharges} and the commutator \eqref{com}, we are now ready to compute their algebra at linear and quadratic order in the same helicity sector. We present the calculations for the positive helicity sector, however similar results hold for the negative one as well. 

\subsection{Linear order}\la{sec:YMlin-alg}

In this section we compute the linear charge algebra in the positive helicity sector. We start by evaluating 
\be 
\begin{split}
    [R_{s+}^2(\tau), R_{s'+}^1(\tau')] &= [R_{s+}^2(\tau), i^{s'}\int_S \Tr\left( D^{s'+1} \tau' \mathscr{F}_+^{\dagger}(s',z) \right)]\\
    &= i^{s'-s} g_{YM}^2\int_S \Tr\left( D^{s'+1}\tau' [P_s(s'+1;\tau), \mathscr{F}_+^{\dagger}(s+s',z)]_{\sG}\right)\\
    &= i^{s'-s} g_{YM}^2 \int_S  \sum_{n = 0}^s \frac{(-s'-1)_n}{n!}
    \Tr\left([D^{s'+1} \tau', D^n \tau ]_{\sG}  D^{s-n} \mathscr{F}_+^{\dagger}(s+s',z) \right),
\end{split}
\ee
where in the first line we used the definition of the linear charge \eqref{YR1}, in the second line we used \eqref{R2Fb}, and in the third line we used \eqref{Ps}. Integrating by parts and using the binomial expansion, we find
\be 
\begin{split}
     [R_{s+}^2(\tau), R_{s'+}^1(\tau')] &= 
     i^{s'+s} g_{YM}^2  \sum_{n = 0}^s \sum_{p = 0}^{s-n}(-1)^{n} \left(\begin{matrix}
         s-n\\
         p
     \end{matrix} \right) \frac{(-s'-1)_n}{n!} \\
     &\times \int_S \Tr\left([ D^{s+s'+1-n-p}\tau', D^{n+p}\tau]_{\sG} \mathscr{F}_+^{\dagger}(s+s',z) \right)\\
     &= i^{s'+s} g_{YM}^2  \sum_{p = 0}^s\left(\begin{matrix}
         s+s'+1\\
         p
     \end{matrix} \right) \int_S \Tr\left([D^{s+s'+1-p} \tau', D^p \tau]_{\sG} \mathscr{F}^{\dagger}_+(s+s',z) \right),
\end{split}
\ee
where in the last line we shifted variables $p \rightarrow p-n$, switched sums $\sum_{n = 0}^s \sum_{p = n}^s = \sum_{p = 0}^{s} \sum_{n = 0}^p$ and evaluated the sum over $n$. 

The linear contribution to the charge commutators is found by adding the term with $s \leftrightarrow s', \tau \leftrightarrow \tau'$, namely
\be 
[R_{s+}(\tau), R_{s'+}(\tau')]^1 = [R_{s+}^2(\tau), R_{s'+}^1(\tau')] - [R_{s'+}^2(\tau'), R_{s+}^1(\tau)].
\ee
This can be immediately evaluated by noting that the binomial coefficient is invariant under $p \rightarrow -p + s+s' + 1$ 
\be 
\left(\begin{matrix}
         s+s'+1\\
         p
     \end{matrix} \right) \rightarrow \left(\begin{matrix}
         s+s'+1\\
         p
     \end{matrix} \right)\,,
\ee
while
\be 
\sum_{p = 0}^{s'} D^{s+s'+1-p} \tau^a D^p \tau'^b = \sum_{p=s+1}^{s+s'+1} D^{s+s'+1-p} \tau'^b D^p \tau^a.
\ee
As a result, it follows that
\be 
[R_{s+}(\tau), R_{s'+}(\tau')]^1 &= i^{s'+s} g_{YM}^2\sum_{p = 0}^{s+s'+1}\left(\begin{matrix}
         s+s'+1\cr
         p
     \end{matrix} \right) \int_S \Tr\left([D^{s+s'+1-p} \tau',D^p \tau]_{\sG} \mathscr{F}^{\dagger}_+(s+s',z) \right)\cr
     &= - g_{YM}^2 R_{s+s',+}^1([\tau, \tau']_{\sG}).
\ee

\subsection{Quadratic order}\la{sec:YMqua-alg}

The quadratic commutator receives two types of contributions, namely
\be 
\begin{split}
[R_{s+}(\tau), R_{s'+}(\tau')]^2 &= [R_{s+}^2(\tau), R_{s'+}^2(\tau')] + [R^1_{s+}(\tau), R_{s'+}^3(\tau')] + [R_{s+}^3(\tau), R_{s'+}^1(\tau')].
\end{split}
\ee
We will show that, quite miraculously, the local contribution to the quadratic-quadratic charge commutator that spoils the algebra is precisely cancelled by the cubic-linear commutators. The remaining pieces of the cubic-linear commutators ensure that the global algebra \eqref{GlobalCharge} is promoted to a local one. 
In the following sections we evaluate the quadratic-quadratic and linear-cubic contributions. We present the main steps leading to the cancellation and defer the details to Appendix \ref{App:YMalg}. 

\subsubsection{Quadratic charge commutator}
\label{sec:quad}

We start by computing the quadratic charge commutator 
\be 
\begin{split}
[R_{s+}^2(\tau), R_{s'+}^2(\tau')] &= \left[R_{s+}^2(\tau), \frac{i^{-s'}}{2\pi}\sum_{\ell = 0}^{\infty} \int_s \Tr\left([P_{s'}(\ell + 1;\tau'), \mathscr{F}^{\dagger}_+(s' + \ell)]_{\sG} \mathscr{A}_+(\ell) \right)\right]\\
&\equiv I_1(s, \tau; s', \tau') + I_2(s, \tau; s', \tau'),
\end{split}
\ee
where $I_1$, $I_2$ arises from the action of $R_{s+}^{2}$ on $\mathscr{F}^{\dagger}_+$ and $\mathscr{A}_+$ using the  charge actions \eqref{R2Fb} and \eqref{R2Ab}.
They read 
\be 
I_1(s, \tau; s', \tau') &= \frac{i^{-s-s'}}{2\pi} g_{YM}^2\sum_{\ell = 0}^{\infty} \int_{S} \Tr\left([P_{s'}(\ell+1;\tau'),[P_s(s' + \ell + 1;\tau), \mathscr{F}^{\dagger}_{+}(s' + s+ \ell)]_{\sG}]_{\sG} \mathscr{A}_+(\ell) \right),
\cr
I_2(s, \tau; s', \tau')&= \frac{i^{s-s'}}{2\pi} g_{YM}^2\sum_{\ell = s}^{\infty} \int_{S} \Tr\left([P_{s'}(\ell+1;\tau'), \mathscr{F}^{\dagger}_{+}(s' + \ell)]_{\sG} [P_s(-\ell - 1;\tau), \mathscr{A}_+(\ell - s)]_{\sG} \right).\nonumber
\ee
The transposition property \eqref{Transpose}  simply implies (after shifting $\ell \rightarrow \ell + s$ in $I_2$) that 
\be 
I_2(s, \tau; s', \tau') = -I_1(s', \tau'; s, \tau).
\ee

It therefore suffices to evaluate $I_1$ and then antisymmetrize in $(s,\tau;s', \tau')$.
We have
\be 
\label{I1}
\begin{split}
I_1(s,\tau; s', \tau') 
&=  \frac{i^{-s-s'}}{2\pi} g_{YM}^2\sum_{\ell = 0}^{\infty} \sum_{n = 0}^{s'} \sum_{m = 0}^s \sum_{p = 0}^{s' - n} \frac{(-\ell - 1)_n}{n!} \frac{(-s' -1 -\ell)_m}{m!} \left(\begin{matrix} s' - n\\
p
\end{matrix} \right) \\
&\times \int_{S} \Tr\left(  [D^n \tau',[D^{p+m} \tau, D^{s+s' -p-n-m}  \mathscr{F}^{\dagger}_{+}(s' + s+ \ell)]_{\mathscr{G}}]_{\mathscr{G}} \mathscr{A}_+(\ell)\right) , \\
\end{split}
\ee
where in the second line we used the binomial expansion. 
We now change variables $p \rightarrow p - m - n$ and perform the sum over $m$ upon changing sums
\be
\sum_{m = 0}^s \sum_{p = m + n}^{s' + m} = \sum_{p = n}^{s + s'} \sum_{m ={\rm max}[ 0,p -s']}^{{\rm min}[p - n, s]}.
\ee
The different cases are worked out in Appendix \ref{App:YMalg-qua}, the result being that the sum splits into two contributions 
\be 
I_1(s, \tau; s', \tau')=J_1(s, \tau; s', \tau')+J'_1(s, \tau; s', \tau')
\ee 
where 
\be 
J_1 =&  \frac{i^{-s-s'}g_{YM}^2}{2\pi} \sum_{\ell = 0}^{\infty} \sum_{n=0}^{s'}  \sum_{p = n}^{s+s'} (-1)^p \left(\begin{matrix} p\\
n
\end{matrix} \right) \left(\begin{matrix} \ell + p\\
\ell
\end{matrix} \right)\cr
&\times\int_{S}\Tr\left( [D^n \tau',[ D^{p-n} \tau, D^{s+s' -p}  \mathscr{F}^{\dagger}_{+}(s' + s+ \ell)]_{\sG}]_{\sG} \mathscr{A}_+(\ell)\right),
\cr
J_1'=& \frac{i^{-s-s'}}{2\pi} g_{YM}^2\sum_{\ell = 0}^{\infty} \sum_{n = 0}^{s'-1} \sum_{p = n+s+1}^{s' + s}  (-1)^{s+n} \left(\begin{matrix} \ell + n\\
n \end{matrix} \right)\left(\begin{matrix} 1 + \ell + s + s'\\
s + 1 \end{matrix} \right)\left(\begin{matrix} s'- n\\
p - n - s - 1\end{matrix} \right) \cr
&\times  {F}_1(s,s') \int_{S}\Tr\left( [D^n \tau',[ D^{p-n} \tau, D^{s+s' -p}  \mathscr{F}^{\dagger}_{+}(s' + s+ \ell)]_{\sG}]_{\sG} \mathscr{A}_+(\ell)\right) .\label{hypergeo}
\ee 
In \eqref{hypergeo} we defined the  hypergeometric function
\be 
\label{eq:F1}
{F}_1(s, s') \equiv  {}_3F_2\left[1, 1 + n - p + s, 2 + \ell + s  + s'; s + 2, 2 - p + s + s';1 \right].
\ee
The  term $J_1$ is present for all ranges of admissible $p$, while the second term $J'_1$ arises only for $p > n + s$. This means that the  term  $J'_1$ vanishes for global transformation parameters.
We conclude that the commutator is 
\be 
[R_{s+}^2(\tau), R_{s'+}^2(\tau')] = J_1(s,\tau; s', \tau')+J_1'(s,\tau; s', \tau') - (s \leftrightarrow s', \tau \leftrightarrow \tau').
\ee

\subsubsection{Cubic charge commutators}\la{sec:ccc}

 Next we compute the cubic-linear charge commutator
 \be 
 \label{cubic-commi3i4}
 \begin{split}
 [R^1_{s+}(\tau), R_{s'+}^3(\tau')] 
:= I_3(s, \tau;s', \tau') + I_4(s, \tau; s', \tau'),
 \end{split}
  \ee
  where $I_3$ and $I_4$ are respectively associated with the commutator of $R^1_{s+}$ in \eqref{YR1} with $A_+(\ell)$ and $A_+(m)$ in \eqref{YR3}.
 For the first contribution, using that $[R_{s+}^1(\tau), \mathscr{A}_+(\ell)]=-i^s 2\pi g_{YM}^2 \delta_{s,\ell}  D^{s+1}\tau$, we find 
  \be 
 \label{eq:I3}
 \begin{split}
 I_3(s,\tau;s',\tau') &= \frac{i^{-s'+s}}{2\pi} g_{YM}^2  \sum_{m = 0}^{\infty} \sum_{n = 0}^{s'-1} \sum_{k = n}^{s' - 1} \sum_{p = 0}^{k - n} (-1)^k \left(\begin{matrix}
 s + n\\
 s
 \end{matrix} \right)\left(\begin{matrix}
 s + m+ k+1\\
 m
 \end{matrix} \right) \left(\begin{matrix}
 k - n\\
 p
 \end{matrix} \right)\\
 &\times \int_S \Tr\left( [D^{n + p} \tau', D^{s + 1 + k - n - p} \tau]_{\sG} [\mathscr{A}(m), D^{s' - k - 1} \mathscr{F}^{\dagger }(s + s' + m)]_{\sG} \right).
 \end{split}
 \ee
 After a straightforward series of changes of variables and sum switches that we detail in Appendix \ref{app:cubic}, this can be shown to simplify to 
 \be 
 \label{I3-final}
 \begin{split}
 I_3{(s,\tau;s',\tau')} &= \frac{i^{{-}(s+s')}}{2\pi} g_{YM}^2  \sum_{m = 0}^{\infty}  \sum_{n = 0}^{s'{-1}} \sum_{k = n+ s + 1}^{s + s'}  (-1)^{k} \left(\begin{matrix}
 k\\
 n
 \end{matrix} \right)\left(\begin{matrix}
 m+ k\\
 m
 \end{matrix} \right)\\
 &\times \int_S \Tr \left( [D^{n} \tau', D^{ k - n} \tau]_{\sG} [ D^{s + s' - k} \mathscr{F}^{\dagger}(s + s' + m),\mathscr{A}(m)]_{\sG} \right).
 \end{split}
 \ee
For the second contribution, direct binomial expansion yields 
 \be 
 \label{I4-in}
 \begin{split}
 I_4(s, \tau;s', \tau') &= \frac{i^{s - s'}}{2\pi} g_{YM}^2  \sum_{\ell = 0}^{\infty} \sum_{n = 0}^{s'-1} \sum_{k = n}^{s' - 1} (-1)^n \left(\begin{matrix} 
 \ell + n\\
 \ell
 \end{matrix}\right) \left(\begin{matrix} 
 s + k + \ell +1\\
 s
 \end{matrix}\right)  \\
 & \times \int_S \Tr\left[[D^n\tau', \mathscr{A}_+(\ell)]_\sG D^{k - n} \Big( [D^{s + 1} \tau, D^{s' - k - 1} \mathscr{F}^{\dagger}_+(s + \ell + s')]_\sG \Big) \right]\\
 &= \frac{i^{s - s'}}{2\pi} g_{YM}^2  \sum_{\ell = 0}^{\infty} \sum_{n = 0}^{s'-1} \sum_{k = n}^{s' - 1} (-1)^n \left(\begin{matrix} 
 \ell + n\\
 \ell
 \end{matrix}\right) \left(\begin{matrix} 
 s + k + \ell +1\\
 s
 \end{matrix}\right)  \\
 & \times \sum_{p = 0}^{k - n} \left(\begin{matrix} 
 k - n\\
 p
 \end{matrix} \right) \int_S \Tr\left([D^n\tau', \mathscr{A}_+(\ell)]_\sG [ D^{s+1+p}\tau, D^{s' - n -p - 1} \mathscr{F}^{\dagger}_+(s + \ell + s')]_\sG \Big) \right).\\
 \end{split}
 \ee
 After a short series of straightforward manipulations detailed in Appendix \ref{app:cubic2} we find 
 \be \label{simp4}
 I_4(s,\tau; s', \tau') = J_4(s,\tau; s', \tau') + J_4'(s,\tau; s', \tau'),
 \ee
 where 
 \be \label{simp41}
 J_4{ ( s, \tau;s', \tau')} &= - \frac{i^{-s' - s}}{2\pi} g_{YM}^2 \sum_{\ell = 0}^{\infty} \sum_{n=0}^{s'-1}\sum_{p=n+s+1}^{s+s'} (-1)^p \left(\begin{matrix} 
 \ell + p\\
 \ell
 \end{matrix}\right) \left(\begin{matrix} 
 p\\
 n
 \end{matrix}\right)\cr
 &\times \int_S \Tr\left(   [D^n\tau',[D^{p - n} \tau, D^{s + s' - p} \mathscr{F}^{\dagger}_+(s  + s'+ \ell)]_{\sG}]_{\sG} \mathscr{A}_+(\ell)\right)
 \ee
 and 
 \be \label{simp42}
 J_4' ( s, \tau;s', \tau') &= \frac{i^{-s' - s}}{2\pi} g_{YM}^2  \sum_{\ell = 0}^{\infty} \sum_{n=0}^{s'-1}\sum_{p=n+s+1}^{s+s'} (-1)^{n+s} \left(\begin{matrix} 
 \ell + n\\
 \ell
 \end{matrix}\right) \left(\begin{matrix} 
 s' - n\\
 p - s - 1 - n
 \end{matrix}\right) \left(\begin{matrix} 
 s' +s + \ell +1\\
 s
 \end{matrix}\right) \cr
 &\times  F_2(s,s') \int_S \Tr\left(   [D^n\tau',[D^{p - n} \tau, D^{s + s' - p} \mathscr{F}^{\dagger}_+(s  + s'+ \ell)]_{\sG}]_{\sG} \mathscr{A}_+(\ell)\right).
 \ee
Here we defined
\be 
 F_2(s, s') \equiv {}_3F_2[1, 1 - n + s', 2 + l + s + s', 2 + \ell + s', 
   2 - p + s + s', 1].
 \ee
From this we conclude that 
\be 
 \begin{split}
 [R^1_{s+}(\tau), R_{s'+}^3(\tau')] 
 \equiv I_3(s, \tau;s', \tau') + J_4(s, \tau; s', \tau')+ J'_4(s, \tau; s', \tau').
 \end{split}
  \ee
{From \eqref{I3-final} and \eqref{I4-in} it is easy to see that \eqref{cubic-commi3i4} vanishes provided that $\tau, \tau'$ obey the global charge condition \eqref{YMGcon}.}

\subsubsection{Full commutators}\la{sec:fcc}

We can now put everything together. We first notice that, quite remarkably, the contributions from the hypergeometric functions cancel each other! In particular we find that
\be 
J_1'(s,\tau;s',\tau') +  J_4'(s, \tau; s', \tau') = 0
\ee
due to the hypergeometric identity 
\be 
F_1(s,s')=- \frac{(s + 1)}{(1+ s' + \ell)} F_2(s,s')
\ee 
proven in  Appendix \ref{app:everything}. It then follows that
 \be 
 \begin{split}
 &  
  J_1(s, \tau; s', \tau')  + J_4(s,\tau;s',\tau') 
 \\
 &= \frac{i^{-s' - s}}{2\pi} g_{YM}^2 \sum_{\ell = 0}^{\infty}\sum_{n = 0}^{ {s'}} \sum_{p = n}^{s + n}  (-1)^p \left(\begin{matrix} 
 \ell + p\\
 \ell
 \end{matrix}\right) \left(\begin{matrix} 
 p\\
 n
 \end{matrix}\right) \\
 &\times \int_S \Tr \left([D^n\tau',   [D^{p - n} \tau, D^{s + s' - p} \mathscr{F}^{\dagger}_+(s + \ell + s') ]_{\sG}]_{\sG}\mathscr{A}_+(\ell)\right).\\
 \end{split}
 \ee
 This  equality follows from the cancellations of sums 
 \be
  \sum_{n = 0}^{s' } \sum_{p = n}^{s+s'}- 
  \sum_{n = 0}^{s'-1 } \sum_{p = n+ s + 1}^{s + s'} 
 = \sum_{n = 0}^{s' } \sum_{p = n}^{n+s}.
 \ee
 In  Appendix  \ref{app:everything} we show as well that the anti-symmetrization of $J_1+J_4$ under the exchange $(s,\tau)\leftrightarrow (s',\tau')$ simplifies into 
 \be 
 & J_1(s, \tau; s', \tau')  + J_4(s,\tau;s',\tau') 
 - (s \leftrightarrow s', \tau \leftrightarrow \tau') 
 \cr
 &= \frac{i^{-s' - s}}{2\pi} g_{YM}^2 \sum_{\ell = 0}^{\infty}\sum_{n = 0}^{ s'} \sum_{p = n}^{s + n}  (-1)^p \left(\begin{matrix} 
 \ell + p\\
 \ell
 \end{matrix}\right) \left(\begin{matrix} 
 p\\
 n
 \end{matrix}\right)
 \cr
 &\times \int_S \Tr \left([[D^n\tau',   D^{p - n} \tau]_{\sG}, D^{s + s' - p} \mathscr{F}^{\dagger}_+(s + \ell + s') ]_{\sG}\mathscr{A}_+(\ell)\right).
 \ee 
 
 To evaluate the commutator at quadratic order we have to add  this contribution to $I_3(s,\tau;s'\tau')- I_3(s',\tau';s,\tau)$ given by \eqref{I3-final}. 
 As a result, we find as shown in  Appendix  \ref{app:everything} that 
 \be 
\begin{split}
[R_{s+}(\tau), R_{s'+}(\tau')]^2 =&   \frac{i^{-s-s'}}{2\pi} g_{YM}^2\sum_{\ell = 0}^{\infty}
\left(  \sum_{n = 0}^{s'}  \sum_{p = n}^{s+n} +   \sum_{p = s + 1}^{s + s'}  \sum_{n = 0}^{p -s-1} +  \sum_{p = s' + 1}^{s + s'} \sum_{n = 1+s'}^{p}  \right)(-1)^p \left(\begin{matrix} p\\
n
\end{matrix} \right) \left(\begin{matrix} \ell + p\\
\ell
\end{matrix} \right) \\
&\times \int_{S} \Tr\left( [D^n \tau', D^{p-n} \tau]_\sG [D^{s+s' -p}  \mathscr{F}^{\dagger}_{+}(s' + s+ \ell), \mathscr{A}_+(\ell)]_\sG \right). \label{R22}
\end{split}
\ee
After a series of straightforward manipulations  described in Appendix \ref{app:everything}, the sum can be simply repackaged as $\sum_{p=0}^{s+s'}
  \sum_{n=0}^{p}$. The sum over $n$ can be reabsorbed into $D^p[\tau,\tau']$ using the Leibniz rule and we remarkably find that \eqref{R22} reduces to 
\be
\label{eq:final-YM}
\boxed{
[R_{s+}(\tau) ,R_{s'+}(\tau') ]^2=
-g_{\rm{YM}}^2 R^{2}_{s+s',+}([\tau,\tau']_{\sG}).
}
\ee

\section{Gravity corner charges}\la{sec:GR}

Similarly to the YM case, in gravity the vacuum asymptotic Einstein's equations (EE) around null infinity can be recast as  a set of recursive differential equations for  higher spin gravitational charge aspects $\mathcal{Q}_s$ given by \cite{Freidel:2021qpz, Freidel:2021dfs, Freidel:2021ytz}
\be \la{GR-rec}
\p_u\mathcal{Q}_s = D \mathcal{Q}_{s - 1} + \frac{s + 1}{2} C Q_{s - 2}, \quad \mathcal{Q}_{-2} = \frac{1}{2} \p_u N, \quad \cQ_{-1}= \frac12 D N, \quad s\geq0,
\ee
with  $C(u,z)$ representing the shear field encoding radiation data, $N=\p_u C^*$ representing the news field and $ \mathcal{Q}_0$ the Bondi mass. The relation between \eqref{GR-rec} and the vacuum EE is exact up to $s=3$ \cite{Newman:1968uj}; for $s\geq 4$ corrections in higher powers of the shear field are expected to appear. Initially neglecting those corrections---that do not affect the linear order same helicity algebra---it was shown in  \cite{Freidel:2021ytz} that the dynamical system defined by \eqref{GR-rec} provides a representation of the $w_{1+\infty}$ loop algebra at linear order. This established a direct relation between the 
 celestial OPE \cite{Fan:2019emx, Pate:2019lpp, Guevara:2021tvr, Guevara:2021abz} 
of two conformal primary gravitons in the collinear and soft limit with the commutator action of the quadratic order (hard) charge contribution on the shear field. This clarified the gravitational origin of the  $w_{1+\infty}$  symmetry originally discovered through celestial OPE techniques  in  \cite{Guevara:2021abz, Strominger:2021mtt}. 

In this second part of the paper, we are going to employ the newly introduced discrete basis for celestial holography \cite{Freidel:2022skz} in order to investigate the fate of such symmetry beyond the linear level.

\subsection{Discrete basis and phase space}\la{GR-db}

We introduce the shear decomposition
 \be\la{C}
C(u) = C_+ (u) + C_-^*(u),
\ee  
with  the   positive and negative helicity graviton components
\begin{subequations}\la{C+-}
\be
C_+(u) := \frac1{2\pi} \int_0^\infty \rd\omega   e^{-i\omega u} \tC_+(\omega)
=-\frac1{2i\pi}  \int_{-\infty}^{+\infty} \rd u' \frac{C(u')}{(u'-u+i\epsilon)} \,,\la{C+}\\
 C_-(u) := \frac1{2\pi} \int_0^\infty \rd\omega   e^{-i\omega u} \tC_-(\omega)
=-\frac1{2i\pi}  \int_{-\infty}^{+\infty} \rd u' \frac{C^{*}(u')}{(u'-u+i\epsilon)}\la{C-}.\ee 
\end{subequations}
Their Fourier and Mellin
 transforms are respectively given by
 \be
\label{Fourier}
\widetilde{C}_+(\omega) &:= \int_{-\infty}^{+\infty} \rd ue^{i\omega u} C(u),
\qquad
\widetilde{C}_-(\omega) := \int_{-\infty}^{+\infty} \rd ue^{i\omega u} C^*(u),
\ee 
and
 \be 
\hC_+(\Delta)  
=  i^{\Delta} \Gamma(\Delta) \int_{-\infty}^{+\infty} \rd u  \frac{ C(u)}{(u{+i\epsilon})^{\Delta}},
\qquad 
\hC_-(\Delta)  
=  i^{\Delta} \Gamma(\Delta) \int_{-\infty}^{+\infty} \rd u  \frac{ C^*(u)}{(u{+i\epsilon})^{\Delta}}\label{hCu}.
\ee
Similarly, we can decompose the news field $N(u)= N_-(u)+N_+^*(u)$, where $N_\pm(u) : =\pa_u C_\pm(u)$.

The  the higher spin, positive and negative energy memory observables are then defined as
\be\la{Mn}
 \mathscr{M}_\pm(n) := \mathrm{Res}_{\Delta = -n} \widehat{N}_\pm(\Delta), \quad n \in \mathbb{Z}_+.
 \ee
These can be conveniently written also as

\be \label{softC}
 \mathscr{M}_+(n) 
 : = \lim_{\omega \rightarrow 0^+} \frac{i^n}{n!}\left( \int_{-\infty}^{+\infty} \rd u\,e^{i\omega u} u^n \pa_uC(u) \right),
   \quad
  \mathscr{M}_-(n) 
 : = \lim_{\omega \rightarrow 0^+} \frac{i^n}{n!}\left( \int_{-\infty}^{+\infty} \rd u\,e^{i\omega u} u^n \pa_uC^*(u) \right) .
 \ee
Note that, in analogy with the YM case, we also have \footnote{{Our conventions here differ by $1/2$ from \cite{Freidel:2022skz}.}}
\be 
\frac{1}{s!}\int_{-\infty}^{\infty} du u^s N(u) = \mathscr{M}^*(s) = \left(i^{-s} \mathscr{M}_-(s) + i^{s} \mathscr{M}_+^*(s) \right) .
\ee

  The memory observables   $\mathscr{M}_\pm(n)$ provide a 
Taylor expansion coefficients of $\widetilde{N}_\pm(\omega)$ around $\omega=0$, as
 \be
 \label{NFourier}
 \widetilde{N}_\pm(\omega)=\sum_{n=0}^\infty  \mathscr{M}_\pm(n) \omega^n, \qquad 
 \mathscr{M}_\pm(n) = \frac{1 }{n!} \left.
 \pa^n_\omega \tN_\pm(\omega)\right|_{\omega=0^+} .
 \ee

 On the other hand,  by evaluating the news Mellin transform at positive integer  conformal dimension  $\Delta= n$, we obtain the Goldstone operators
 \be 
\mathscr{S}_{\pm}(n) :=\lim_{\Delta \rightarrow n} \widehat{N}_{\pm}(\Delta), \quad n \in \mathbb{Z}_+\,.
\ee
The Goldstone modes provide a Taylor expansion of $C_\pm(u)$ around $u=0$, as
\be\label{uexp}
C_\pm (u)
=\f{i}{2\pi } \sum_{n=0}^\infty \frac{(-iu)^n}{n!}   \mathscr{S}_\pm(n)\,.
\ee

The gravitational phase space at asymptotic infinity  is characterized by the radiative symplectic potential  \cite{Ashtekar:1978zz, Ashtekar:1981sf, Ashtekar:2018lor} 
\be
\Theta^{\va \rm{GR}}= \f2\k\int_{\scri^+} N(u,z) \delta C(u,z),
\ee
with $\kappa=\sqrt{32\pi G}$. This can be decomposed (up to a canonical transformation) into  positive and negative helicity components $\Theta^{\va \rm{GR}}=\Theta^{\va \rm{GR}}_++\Theta^{\va \rm{GR}}_-$,  each parametrized by the respective infinite tower  of memory and their conjugate (complex conjugate)  Goldstone operators as
\be \label{Spot}
\Theta^{\va \rm{GR}}_\pm &=  \f2\k\int_{\scri^+}  N_\pm(u,z) \delta C_\pm^*(u,z)
=   \f{1}{i\pi\k } \sum_{n=0}^\infty  \int_S \mathscr{M}_\pm(n,z) \d \mathscr{S}^*_\pm(n,z) \,.
\ee
At the quantum level then, the only non-trivial commutators are 
 \be\la{corn-comm}
[\mathscr{M}_\pm(n,z ),\mathscr{S}_\pm^\dagger(m, z')]= \pi \k   \delta_{n,m}\d^2(z,z').
\ee

\subsection{Charge aspects}\la{GR-hsc}

The charge aspects solving \eqref{GR-rec} can again be expanded  in powers of radiation fields as
\be\cQ_s =\sum_{k=1}^{\max[2,s+1]} \cQ_s^{k}\,.
\la{charge-kexp}
\ee
At a given order $k$ in powers of radiation fields, the renormalized aspects
can be expressed as
\be\la{qren}
\hat q^k_s(u,z)
&=\sum_{n=0}^{s} \frac{(-u)^{s-n}}{(s-n)!} D^{s-n} \cQ^k_n(u,z)\cr
&=\frac12 \p_u^{-1} \sum_{\ell=0}^{s}  (\ell+1)  \frac{(-u)^{s-\ell}}{(s-\ell)!} 
 D^{s-\ell}  \left[C (u,z) \cQ^{k-1}_{\ell-2}(u,z)  \right].
 \ee
 As clear from the expression above, the higher spin charges can be recursively expressed as a nested product of  integrals over $\scri$.  The discrete basis introduced in  \cite{Freidel:2022skz} allows one to eliminate all the time integrals and obtain expressions for the charges as a single integral over a corner at arbitrary value of retarded time $u$. In the rest of the paper we will concentrate on the case $u=0$, but formulas for generalization to arbitrary $u=u_0$ can be found in \cite{Freidel:2022skz}. Let us first provide a brief review of the main ingredients of the new discrete basis.

\subsection{Charges}\la{sec:GR-charges}

As shown in \cite{Freidel:2021ytz, Freidel:2022skz}, the renormalized higher spin  charges in gravity are defined as 
\be\la{Qstau}
Q_s(\tau) := \int_S  q_s(z) \tau(z),
\ee
where
\be\la{Qren2}
q_s(z)=\lim_{u\to-\infty} \hat q_s(u,z).
\ee 
By performing a decomposition into positive and negative helicity components, for each order we can define
\be\la{Qk+-}
 Q^k_s(\tau) =
  -\frac1{2}\left[ i^{{-}s} Q^k_{s+}(\tau)  + i^{s}Q_{s-}^{*k}(\tau)\right].
\ee
The expressions of the positive and negative helicity parts
in terms of the  memory and Goldstone variables
$ \mathscr{M}(n), \mathscr{S}(n)$ for the linear and quadratic orders were derived in  \cite{Freidel:2022skz} and are respectively given by
\be
 Q^1_{s+}(\tau)
 = (-1)^s \int_S   {D}^{s+2}\tau(z) \mathscr{M}^{*}_+(s,z)\,,
 \qquad 
 Q^1_{s-}(\tau)
 = (-1)^s \int_S   \bar D^{s+2}\tau(z) \mathscr{M}^{*}_-(s,z)
 \,,\la{Q1}
\ee
and 
\be
Q^{2 }_{s+}(\tau)
 &=- \frac{  1}{{4}\pi}  \sum_{n=0}^\infty \sum_{\ell=0}^s (-)^{\ell+s} (\ell+1)
\f{(s+n-\ell)_{s-\ell}}{(s-\ell)!}
 \int_S   D^{s-\ell}\tau(z) \mathscr{S}_+(n,z)D^{\ell}  {\mathscr{M}}^*_+(s+n-1,z),
 \la{Q2+}\\
 Q^{2 *}_{s-}(\tau)
 &=- \frac{  1}{{4}\pi}  \sum_{n=0}^\infty \sum_{\ell=0}^s (-)^{\ell+s} (\ell+1)
\f{(s+n-\ell)_{s-\ell}}{(s-\ell)!}
 \int_S   D^{s-\ell}\tau(z) \mathscr{S}^{ *}_-(n,z)D^{\ell}  {\mathscr{M}}_-(s+n-1,z).\la{Q2-}
 \ee
 We also computed the action
 \be\la{Q2Nn}
[ Q^{2}_{s+}(\tau), \mathscr{S}_+(n,z) ]&=-
\frac{(-)^{s} \k}{4}    \sum_{k=0}^s 
 (s-k+1 ) 
 \left(\begin{matrix}
n+3\\
k
\end{matrix}\right)
  D^{k} \tau(z)
  D^{s-k}\mathscr{S}_+(n-s+1,z)\,. 
\ee

In Appendix \ref{App:GRcharges} we compute the cubic charges in the discrete basis and their action on the Goldstone operators. These are given respectively by
\be
Q_{s+}^{3}(\tau)
&=\frac{(-)^{s}}{4 (2\pi)^2} \sum_{n=0}^{\infty} \sum_{k=0}^{\infty} \sum_{\ell=2}^s \sum_{m=0}^{\ell-2}(-)^{\ell} (\ell+1) (m+1)
  \left(\begin{matrix}
s+n-\ell\\
n
\end{matrix}\right)
 \left(\begin{matrix}
s+n+k-m-1\\
k
\end{matrix}\right)
\cr
&\times 
\int_S \rd^2 z \sqrt{q}\, D^{s-\ell} \tau(z)\mathscr{S}_+(n, z) 
\left[
  D^{\ell-m-2} \left[  \mathscr{S}_+(k, z) D^{m}  \mathscr{M}^*_+(s+n+k-2,z) \right]
  \right] \,,\la{Q3}
\ee
and
\be
&[ Q^{3}_{s+}(\tau),  \mathscr{S}_+(n,z) ]= 
\frac{(-)^{s+1}\k}{16\pi}   \sum_{k=0}^{\infty} \sum_{\ell=2}^{s} \sum_{m=0}^{\ell-2}\sum_{p=0}^{\ell-m-2}
  (\ell+1) (m+1)
  \cr
  &\times
  \left(\begin{matrix}
n-k+2-\ell\\
s-\ell
\end{matrix}\right)
 \left(\begin{matrix}
n-m+1\\
k
\end{matrix}\right)
 \left(\begin{matrix}
\ell-m-2\\
p
\end{matrix}\right)
\cr
&\times 
D^{m}\left[
D^{s-m-2-p} \tau(z)  D^p\mathscr{S}_+(n-k-s +2, z) 
\mathscr{S}_+(k, z)  
\right]
 \,.\la{Q3S}
\ee

\section{Gravity corner algebra}\la{sec:GR-alg}

As an application of the new discrete basis, we can verify that the expressions \eqref{Q1}, \eqref{Q2+}, \eqref{Q2-} for the linear and quadratic charges in terms of the memory observables and the  Goldstone modes reproduce the $Lw_{1+\infty}$ symmetry (loop) algebra at linear order, as previously computed in \cite{Freidel:2021ytz}. 
Furthermore, we will exploit the computational advantages of the new basis to prove the validity of the $Lw_{1+\infty}$ loop algebra also at quadratic order, when restricting to wedge sector, and in the general case of local charges for the choice of spins $s=s'=2$ . 

More precisely, the wedge subalgebra $WLw_{1+\infty}\subset Lw_{1+\infty}$ is characterized by the following restriction of the transformation parameters
\be\la{global-GR}
D^{s+2} \tau_s(z)=0\,.
\ee
Note that the linear (soft) charges vanish for this choice of parameters. While on the plane, solutions to \eqref{global-GR} are polynomials of degree $s + 1$ in $z$, on the sphere \eqref{global-GR} admits no non-trivial global solutions. Instead, \eqref{global-GR} can only hold away from points $z_s$ where  $D^{s+2} \tau_s(z)= D^p \delta(z - z_s)$.
The corresponding charge aspects are associated to the  {\it global} components (in a spherical harmonic decomposition) $\Psi_0^{(s-2)}$ in the asymptotic expansion of the $\Psi_0$ Weyl scalar (see \cite{ Freidel:2021ytz} for more details on this relation). These also represent the relevant symmetry sector of the twistor formulation of self-dual
gravity \cite{Adamo:2021lrv}.

\subsection{Linear order commutator}

As shown in Appendix \ref{App:GRalg-lin}, the new basis considerably simplifies the calculations and, by means of the commutation relations \eqref{corn-comm},
 it allows us to recover the $Lw_{1+\infty}$ loop algebra for the positive helicity piece of the charges \eqref{Qk+-}. Explicitly, the  commutator at linear order yields
\be
[Q_{s}(\tau), Q_{s'}(\tau')]_+^1&=[Q^{1}_{s+}(\tau), Q^{2}_{s'+}(\tau')]+[Q^{2}_{s+}(\tau), Q^{1}_{s'+}(\tau')]\cr
&=
 \f{\kappa^2}{{4}}\left[(s'+1) Q^{1}_{s+s'-1,+}(\tau' D \tau) 
-(s+1) Q^{1}_{s+s'-1,+}(\tau D \tau') 
\right]\,.
\ee
The same result holds for the negative helicity piece. Some of the  intricacies for the mixed helicity sector were pointed out in \cite{Freidel:2021ytz}. We expect the computational simplifications brought along by the new basis to help investigate them.

\subsection{Quadratic order   commutator of global charges }

In order to  compute the quadratic order of the bracket $[Q_s(\tau),  Q_{s'}(\tau')]$ in the global sector (denoted  by the subscript $G$), we use the Jacobi relation
\be\la{Jacobi-glo}
&[Q^{2}_{s+}(\tau) , [ Q^{2}_{s'+}(\tau'),   \mathscr{S}_+(n,z) ]]_G-[Q^{2}_{s'+}(\tau') , [ Q^{2}_{s+}(\tau),   \mathscr{S}_+(n,z) ]]_G
\cr
&= [  [Q^{2}_{s+}(\tau) ,Q^{2}_{s'+}(\tau') ], \mathscr{S}_+(n,z) ]_G.
\ee 

Let us introduce the useful operatorial relation
\be
 D^{\alpha-1} \left[(s+1) \tau D +(s+1 -  \alpha) D\tau\right] D^{s-\alpha}
 =\left[ \sum_{k=0}^{s+1} +\sum_{k=s+2}^\infty \right] (s+1-k) \frac{(\alpha)_{k}}{ k! }   (D^{k} \tau ) D^{s-k}\,,\la{magic}
\ee
 which can be proven in terms of the generalized Leibniz rule \eqref{magicYM}. 
For the global charges \eqref{global-GR}, this reduces to (see Appendix \ref{App:GRalg-qua-g})
\be
 D^{\alpha-1} \left[(s+1) \tau D +(s+1 -  \alpha) D\tau\right] D^{s-\alpha}
 = \sum_{k=0}^s (s+1-k) \frac{(\alpha)_{k}}{ k! }   (D^{k} \tau ) D^{s-k}\,,\la{magic-g}
\ee
and we have the quadratic charge global action
\be
[ Q^{2}_{s+}(\tau),  \mathscr{S}_+(n,z) ]_{G}&=-
\frac{(-)^{s} \k}{{4}}  
 D^{n+2} \left[(s+1) \tau D +(s -  n-2) D\tau\right] D^{s-n-3}\mathscr{S}_+(n-s+1,z).\la{R1}
 \ee
From this we obtain (see Appendix \ref{App:GRalg-qua-g})
{\be
&[Q^{2}_{s'+}(\tau') , [ Q^{2}_{s+}(\tau),   \mathscr{S}_+(n,z) ]]_G-[Q^{2}_{s+}(\tau) , [ Q^{2}_{s'+}(\tau'),   \mathscr{S}_+(n,z)]]_G \cr
&=\k\frac{(-)^{s+s'} }{16}
   \sum_{k=0}^{n+3} 
 (s+s'-k) \left(\begin{matrix}
n+3\\
k
\end{matrix}\right)
D^k  [\tau,\tau']  D^{s+s'-1-k}  \mathscr{S}_+(n-s-s'+2,z) \,,\la{Q2Q2}
\ee}
where
\be
[\tau,\tau'] := (s+1) \tau D\tau' - (s'+1) \tau' D\tau\,.
\ee

It can easily be checked (see again Appendix \ref{App:GRalg-qua-g}) that this matches exactly the action
\be
- \frac {\k}{{4}}\left[(s'+1)[ Q^{2}_{s+s'-1,+}(\tau' D \tau) , \mathscr{S}_+(n,z) ]-(s+1) [Q^{2}_{s+s'-1,+}(\tau D \tau') , \mathscr{S}_+(n,z) ]
\right]\,,
\ee
where notice that in this case we {\it do not} need to restrict to the global charges.
Hence, from the Jacobi relation
\eqref{Jacobi-glo}, we immediately obtain
the quadratic commutator of the global charges
\be\la{Q2Q2g}
[Q^{2}_{s+}(\tau), Q^{2}_{s'+}(\tau')]_G=
\frac {\k }{{4}}\left[(s'+1) Q^{2}_{s+s'-1,+}(\tau' D \tau) 
-(s+1) Q^{2}_{s+s'-1,+}(\tau D \tau') 
\right]\,.
\ee

\subsection{Quadratic order commutator of the local charges $s=s'=2$}

As a final step of this paper towards the full proof of the validity of the local $Lw_{1+\infty}$ algebra, we show it here for the simpler case $s=s'=2$. We begin with general considerations and then specialize to this restriction on the spins.

Consider the Jacobi identity
\be
[Q_{s+}(\tau) , [Q_{s'+}(\tau'), \mathscr{S}_+(n,z) ]]-[Q_{s'+}(\tau') , [ Q_{s+}(\tau), \mathscr{S}_+(n,z) ]] = [[Q_{s+}(\tau) ,Q_{s'+}(\tau') ],\mathscr{S}_+(n,z) ],
\ee
which at quadratic order gives 
\be\la{Jacobi-loc}
&[Q^{2}_{s+}(\tau) , [Q^{2}_{s'+}(\tau'), \mathscr{S}_+(n,z) ]]-[Q^{2}_{s'+}(\tau') , [ Q^{2}_{s+}(\tau), \mathscr{S}_+(n,z) ]] \cr
&
+[Q^{ 1}_{s+}(\tau) , [Q^{3}_{s'+}(\tau'), \mathscr{S}_+(n,z) ]]-[Q^{1}_{s'+}(\tau') , [ Q^{3}_{s+}(\tau), \mathscr{S}_+(n,z) ]] 
\cr
&= [[Q_{s+}(\tau) ,Q_{s'+}(\tau') ]^{(2)},\mathscr{S}_+(n,z) ]\,, \cr
\ee
where we used the fact that
\be
[Q^{3}_{s'+}(\tau') , [ Q^{ 1}_{s+}(\tau),  \mathscr{S}_+(n,z) ]]=0,
\ee
and
\be
[Q_{s+}(\tau) ,Q_{s'+}(\tau') ]^{(2)}=
[Q^{2}_{s+}(\tau) ,Q^{2}_{s'+}(\tau') ]
+[Q^{ 1}_{s+}(\tau) ,Q^{3}_{s'+}(\tau') ]
-[ Q^{1}_{s'+}(\tau') ,Q^{3}_{s+}(\tau)].
\ee

As shown in the derivation of \eqref{Q2Q2g}, the restriction \eqref{global-GR} to the wedge sector of the nested commutators involving quadratic charges  in \eqref{Jacobi-loc} is sufficient to yield the desired result. Therefore, the goal is to show that the remaining contributions (the `remainders') to those commutators are cancelled exactly by those on the LHS of \eqref{Jacobi-loc} involving linear and cubic charges.

The remainder nested commutators are of two types
\be\la{Rem}
[Q^{2}_{s'+}(\tau') , [ Q^{2}_{s+}(\tau),  \mathscr{S}_+(n,z) ]_{R}]\,,\quad 
[Q^{2}_{s'+}(\tau') , [ Q^{2}_{s+}(\tau),  \mathscr{S}_+(n,z) ]_G]_{R},
\ee
where we have defined
\be
[ Q^{2}_{s+}(\tau),  \mathscr{S}_+(n,z) ]_{R}&:=
\frac{(-)^{s} \k}{{4}}   \sum_{k=s+1}^\infty 
 (s-k+1 ) 
 \left(\begin{matrix}
n+3\\
k
\end{matrix}\right)
  D^{k} \tau(z)
  D^{s-k}\mathscr{S}_+(n-s+1,z),\la{R2grav}
\ee
which is the complementary contribution to the action \eqref{Q2Nn} when the global condition \eqref{global-GR} is relaxed (see derivation of \eqref{R1} in Appendix \ref{App:GRalg-qua-g}).

The two remainder contributions \eqref{Rem} are computed in Appendix \ref{App:algebra-quadratic-22} and they are respectively
\be\la{R2try3}
&[Q^{2}_{s'+}(\tau') , [ Q^{2}_{s+}(\tau),  \mathscr{S}_+(n,z) ]_{R}]
\cr
&=-\frac{(-)^{s+s'} \kappa^4}{{16}}    \sum_{k=s+1}^\infty
   \sum_{\ell=0}^{s'}  \sum_{m=0}^{\infty}
   (s-k+1 ) 
 \left(\begin{matrix}
n+3\\
k
\end{matrix}\right)
 (s'-\ell+1 ) 
 \left(\begin{matrix}
n-s+4\\
\ell
\end{matrix}\right)
(-)^m \f{(k-s+m-1)!}{m!(k-s-1)!}
\cr
&\times
  D^{k} \tau(z)
D^{\ell+m} \tau'(z)
  D^{s+s'-\ell-k-m}\mathscr{S}_+(n-s-s'+2,z),
\ee
and 
\be\la{R1R2-main}
&[Q^{2}_{s'+}(\tau') , [ Q^{2}_{s+}(\tau),  \mathscr{S}_+(n,z) ]_{G}]_{R}
\cr
&= -\frac{(-)^{s+s'} \kappa^4}{{16}} 
 \sum_{k=s'+1}^{n-s+4}\sum_{m=0}^{n+2}
   (s'-k+1 ) 
 \left(\begin{matrix}
n-s+4\\
k
\end{matrix}\right)
\left(\begin{matrix}
n+2\\
m
\end{matrix}\right)
\cr
&\times
\bigg(
(s+1) D^m \tau  D^{s-m}
 \left[ D^{k} \tau'(z)
  D^{s'-k}\mathscr{S}_+(n-s-s'+2,z)\right]
  \cr
  &
 +(s-n-2) D^{m+1}\tau D^{s-1-m}
 \left[ D^{k} \tau'(z)
  D^{s'-k}\mathscr{S}_+(n-s-s'+2,z)\right]\
    \bigg).
\ee
The remaining nested commutators involving linear and cubic charges can be computed by acting with $ Q^{1}_{s'+}(\tau')$ on \eqref{Q3S}. We don't give the general expression here, but instead we compute it for the case $s=2$, namely (see Appendix \ref{App:algebra-quadratic-22})
\be\la{Q1Q3S}
&[Q^{ 1}_{s'+}(\tau') , [ Q^{3}_{2+}(\tau),  \mathscr{S}_+(n,z) ]]
=
 \frac{3 (-)^{s'}\kappa^4}{16} 
 \left(\begin{matrix}
n+2\\
s'+1
\end{matrix}\right)
\tau_{2}(z) D^{s'+2} \tau'_{s'}(z) \mathscr{S}_+(n-s', z).
\ee
We now specialize the expressions \eqref{R2try3}, \eqref{R1R2-main} to the case $s=s'=2$. We obtain
\be
&[Q^{2}_{2+}(\tau') , [ Q^{2}_{2+}(\tau),  \mathscr{S}_+(n,z) ]_{R}]
+
[Q^{2}_{2+}(\tau') , [ Q^{2}_{2+}(\tau),  \mathscr{S}_+(n,z) ]_{G}]_{R}
\cr
&=
\frac{3\kappa^4}{16}  
\left[
 \left(\begin{matrix}
n+3\\
4
\end{matrix}\right)
 \tau'_{2}(z) D^{4} \tau_2(z)
+ 
 \left(\begin{matrix}
n+2\\
4
\end{matrix}\right)
 \tau_2  
D^{4} \tau'_{2}(z)
\right]\mathscr{S}_+(n-2,z).
\ee
On the other hand, the contribution \eqref{Q1Q3S} for $s = s'=2$ gives
\be\la{Q1Q3S2}
&[Q^{ 1}_{2+}(\tau') , [ Q^{3}_{2+}(\tau),  \mathscr{S}_+(n,z) ]]
=
 \frac{3 \kappa^4}{16} 
 \left(\begin{matrix}
n+2\\
3
\end{matrix}\right)
\tau_{2}(z) D^{4} \tau'_{2}(z) \mathscr{S}_+(n-2, z).
\ee
Therefore, upon antisymmetrization in $\tau\leftrightarrow \tau'$, we see that
\be
&[Q^{2}_{2+}(\tau') , [ Q^{2}_{2+}(\tau),  \mathscr{S}_+(n,z) ]_{R}]
+
[Q^{2}_{2+}(\tau') , [ Q^{2}_{2+}(\tau),  \mathscr{S}_+(n,z) ]_{G}]_{R}
\cr
&+ [Q^{ 1}_{2+}(\tau') , [ Q^{3}_{2+}(\tau),  \mathscr{S}_+(n,z) ]]
-\tau\leftrightarrow \tau' 
=0,
\ee
and hence
\be\la{Q2Q2l}
[Q_{2+}(\tau), Q_{2+}(\tau')]^{(2)}=
[Q^{2}_{2+}(\tau), Q^{2}_{2+}(\tau')]_G=
\frac {3\k  }{4}\left[ Q^{2}_{3+}(\tau' D \tau) 
- Q^{2}_{3+}(\tau D \tau') 
\right]\,.
\ee

\section{Conclusions}\la{sec:conc}

In this paper we extracted a tower of non-linear operators from the asymptotic Yang--Mills equations and demonstrated that they form a representation of a higher-spin symmetry loop algebra on Fock space. This algebra contains a global subalgebra, which at the linear order is the phase space realization of the celestial algebra of soft gluon operators found in \cite{Guevara:2021abz, Strominger:2021mtt}. Remarkably, we found that this algebra admits a local enhancement and continues to hold at quadratic order upon inclusion of new cubic terms as dictated by the equations of motion. The steps leading to this result involved a series of miraculous cancellations, which we believe hints at a deeper connection between symmetry and the asymptotic Yang--Mills equations. It would be great to find an elegant way of deriving the loop algebra \eqref{YM-alg-intro} directly from the recursive towers of non-linear differential equations \eqref{YM-diff-eq-intro}, perhaps by employing or generalizing the methods of \cite{DS}.

From a physical perspective, the  implications of the infinite-dimensional symmetry for scattering remain rather unclear. As a first step in this direction, one should try to understand the signatures of the cubic and higher order components of the charges in scattering amplitudes. Moreover, the full Yang--Mills equations will include further non-linear corrections which deserve a better understanding (see \cite{NicoandI} for progress in this direction). It would be very interesting to understand in what way these corrections, as well as coupling to matter, would affect the symmetry structures found in this work. 

Central to this work was the algebra of quadratic charges, also known as hard charges. On the other hand, the celestial symmetry algebras of \cite{Guevara:2021abz, Strominger:2021mtt} were associated with soft operators. We would like to have a better understanding of the dictionary between symmetry generators in celestial CFT and realization of the symmetry algebras on the bulk Fock space. 

In gravity, the non-linear charges were shown to correspond to the higher multipole moments of the gravitational field and hence directly related to gravitational observables such as the memory effect \cite{Compere:2022zdz, Blanchet:2023pce}. It would be fascinating to explore the role of symmetry in constraining observables of gauge theory and gravity.
We leave this to future work.

\section*{Acknowledgements}

We would like to thank Sabrina Pasterski and Yangrui Hu for many  discussions. 
Research at Perimeter Institute is supported in part by the Government of Canada through the Department of Innovation, Science and Economic Development Canada and by the Province of Ontario through the Ministry of Colleges and Universities. D.P. has received funding from the European Union's Horizon 2020 research and innovation programme under the Marie Sklodowska-Curie grant agreement No. 841923. A.R. is supported by the Heising-Simons Foundation ``Observational Signatures of Quantum
Gravity'' collaboration grant 2021-2817 and acknowledges Perimeter Institute for hospitality while this work was completed.

\appendix

\section{YM charges}\la{App:YMcharges}

From the general expression \eqref{Rk} for the YM  higher spin charge aspects, the quadratic and cubic charge aspects read respectively 
\be\la{R2}
\YM^2_s(u,z)&=i\sum_{n=0}^s  (\p_u^{-1})^{s-n+1} D^{s-n} \left[A_z^{(0)}(u,z) ,\YM^1_{n-1}(u,z)  \right]_\sG\cr
&=i\sum_{n=0}^s  (\p_u^{-1})^{s-n+1} D^{s-n} \left[A_z^{(0)}(u,z) ,(\p_u^{-1}D)^{n}  F_{\bz u}^{(0)}(u,z)  \right]_\sG
\cr
&=i\sum_{n=0}^{s}  (\p_u^{-1})^{n+1} D^{n} \left[A_z^{(0)}(u,z) ,(\p_u^{-1}D)^{s-n}  F_{\bz u}^{(0)}(u,z)  \right]_\sG
\ee
and
\be\la{R3}
\YM^3_s(u,z)&=
i\sum_{n=1}^{s}  (\p_u^{-1})^{s-n+1} D^{s-n} \left[A_z^{(0)}(u,z), \YM^{2}_{n-1}(u,z)  \right]_\sG
\cr
&= -\sum_{n= 1}^{s} \sum_{k=0}^{n-1}  (\p_u^{-1})^{s-n+1} D^{s-n}  \left[ A_z^{(0)}(u,z), (\p_u^{-1})^{k+1} D^{k}  \left[ A_z^{(0)} (u,z),  (\p_u^{-1}D)^{n-k-1}  F_{\bz u}^{(0)}(u,z)\right]_\sG\right]_\sG.\cr
\ee

Let us introduce the Leibniz rule for pseudo-differential calculus 
\be 
\pa_{u}^{-1}\left( \frac{u^k}{k!} A(u)\right)&= \sum_{n=0}^{k} \frac{(-1)_n}{n!}\frac{u^{(k-n)}}{(k-n)!}
(\pa_{u}^{-1 })^{n+1}A(u)\cr
&=(-1)^k\sum_{n=0}^{k}\frac{(-u)^{(k-n)}}{(k-n)!}
(\pa_{u}^{-1 })^{n+1}A(u)\,,\la{Leibn}
\ee
where we used 
\be
(-1)_n=(-1)^n n!\,.
\ee

For $k=1$, the renormalized charges \eqref{rk} can be expressed as
\be
 \ym^1_s(z)&= \lim_{u\to-\infty} \sum_{\ell=0}^{s} \frac{(-)^{s-\ell}u^{s-\ell}  (\p_u^{-1})^{\ell+1}}{(s-\ell)!} D^{s+1}   F_{\bz u}^{(0)}(u,z)\cr
 &=(-)^{s+1} \int_{-\infty}^\infty du  \frac{u^s}{s!}  D^{s+1} F_{\bz u}^{(0)}(u,z)\,,
 \ee
 where in the last line and below we recall that $\p_u^{-1} = \int_{\infty}^u du$.
 For $k\geq 2$, we can rewrite
\be
 \ym^k_s(z)&= \lim_{u\to-\infty} \sum_{\ell=0}^{s} \frac{(-)^{s-\ell}u^{s-\ell}}{(s-\ell)!} D^{s-\ell} \YM^k_\ell(u,z)\cr
 &= \lim_{u\to-\infty}i \sum_{n= k-2}^s  \sum_{\ell=n}^{s}
  \frac{(-u)^{s-\ell}}{(s-\ell)!}  
 (\p_u^{-1})^{\ell-n+1}  D^{s-n}  \left[A_z^{(0)}(u,z), \YM^{k-1}_{n-1}(u,z)  \right]_\sG
\cr
  &= \lim_{u\to-\infty} i \sum_{n= k-2}^s  \sum_{\ell=0}^{s-n}
  \frac{(-u)^{s-\ell-n}}{(s-\ell-n)!}  
 (\p_u^{-1})^{\ell+1}  D^{s-n}  \left[A_z^{(0)}(u,z), \YM^{k-1}_{n-1}(u,z)  \right]_\sG
 \cr
  &= \lim_{u\to-\infty} i \sum_{n= k-2}^s  \p_u^{-1}
  \frac{(-u)^{s-n}}{(s-n)!}  
 D^{s-n}  \left[A_z^{(0)}(u,z), \YM^{k-1}_{n-1}(u,z)  \right]_\sG
\cr
  &= - i \sum_{n= k-2}^s \int_{-\infty}^\infty
  \frac{(-u)^{s-n}}{(s-n)!}  
 D^{s-n}  \left[A_z^{(0)}(u,z), \YM^{k-1}_{n-1}(u,z)  \right]_\sG\,,\la{qsk}
\ee
from which
\be
 \ym^2_s(z)
&= - i  \sum_{n=0}^{s}  \int_{-\infty}^\infty du
 \frac{(-)^{s-n}u^{s-n}}{(s-n)!} D^{s-n} \left[A_z^{(0)}(u,z), (\p_u^{-1})^{n} D^{n} F_{\bz u}^{(0)}(u,z) \right]_\sG,
\ee
and 
\be
\ym^3_s(z)
&= \sum_{n= 1}^{s}  \sum_{k=0}^{n-1}\int_{-\infty}^\infty\rd u
 \frac{(-u)^{s-n}}{(s-n)!}
 D^{s-n}  \left[ A_z^{(0)}(u,z) , (\p_u^{-1})^{k+1} D^k  \left[ A_z^{(0)}(u,z), (\p_u^{-1}D)^{n-k-1}  F_{\bz u}^{(0)}(u,z)\right]_\sG\right]_\sG,
\ee
where we used  \eqref{R2}.

From the general expression \eqref{qsk}, we can also write the recursion relation
\be
 \ym^k_s(z)
&= \lim_{u\to-\infty} i \sum_{\ell=0}^{s-k+2}  \p_u^{-1}
  \f{(-u)^{\ell} }{\ell!} \left[D^{\ell} A_z^{(0)}(u,z),
   \sum_{n=\ell}^{s-k+2}
  \frac{(-u)^{n-\ell}}{(n-\ell)!}  
 D^{n-\ell}\YM^{k-1}_{s-n-1}(u,z)  \right]_\sG
\cr
&= \lim_{u\to-\infty} i \sum_{\ell=0}^{s-k+2}  \p_u^{-1}
  \f{(-u)^{\ell} }{\ell!}\left[ D^{\ell} A_z^{(0)}(u,z),
   \sum_{n=0}^{s-\ell -k+2}
  \frac{(-u)^{n}}{n!}  
 D^{n}\YM^{k-1}_{s-\ell-n-1}(u,z)   \right]_\sG
\cr
  &= \lim_{u\to-\infty} i \sum_{\ell=0}^{s-k+2}  \p_u^{-1}
  \f{(-u)^{\ell} }{\ell!}\left[ D^{\ell} A_z^{(0)}(u,z),
    \ym^{k-1}_{s-\ell-1}(u,z) \right]_\sG
    \cr
  &= \lim_{u\to-\infty}i \sum_{\ell=0}^{s-k+2}  \p_u^{-1}
  \f{(-u)^{\ell} }{\ell!}\left[ D^{\ell} A_z^{(0)}(u,z),
    \ym^{k-1}_{s-\ell-1}(u,z) \right]_\sG,
\ee
where in the first line we applied the binomial expansion and switched the sums over $\ell$ and $n$.

\subsection{Corner charge aspects}\la{App:YMcharges-cor}

Let us first of all list the useful the relations
\be
&u^n\p_u^n=(\widehat\Delta-1)_n\,,\qquad \p_u^nu^n=(\widehat\Delta+n-1)_n\,,\qquad
u^{-n}\p_u^{-n}=(\widehat\Delta+n-1)^{-1}_n\,,\cr
& \p_u (\widehat\Delta+\alpha)_n= (\widehat\Delta+\alpha+1)_n \p_u\,,\qquad
\p_u^{-1} (\widehat\Delta+\alpha)_n= (\widehat\Delta+\alpha-1)_n \p_u^{-1}\,,\cr 
& u(\widehat\Delta+\alpha)_n= (\widehat\Delta+\alpha-1)_n u
\,,\qquad
u(\widehat\Delta+n-1)^{-1}_n=(\widehat\Delta+n-2)^{-1}_n u\,,
\la{uDelta}
\ee
valid $\forall~ n\geq 0\,,\alpha\in \Z$ and where we  defined the operator $\widehat\Delta :=\pa_u u$ and the requirement  of the potential field to be  Schwartzian in order to integrate the $\widehat\Delta$ contributions to zero.

The quadratic charge aspects \eqref{r2} can be expressed as
\be
 \ym^2_s(z)=   \ym^2_{s+}(z)+  \ym^{2*}_{s-}(z)\,,
\ee
with
\be
   \ym^2_{s+}(z)
&=- i  \sum_{n=0}^{s}  \int_{-\infty}^\infty du 
 \frac{(-)^{s-n}u^{s-n}}{(s-n)!} D^{s-n} \left[A_{+}(u,z), (\p_u^{-1})^{n} D^{n}F^*_{ +}(u,z) \right]_\sG
 \cr
 &=-\f{i^{-\ell}}{2\pi}   \sum_{n=0}^{s}  \sum_{\ell=0}^\infty  
 \frac{(-)^{s-n}}{(s-n)! \ell!} D^{s-n} \left[\mathscr{A}_+(\ell,z),  \int_{-\infty}^\infty du u^{s+\ell-n} (\p_u^{-1})^{n} D^{n}F^*_{ +}(u,z) \right]_\sG
 \cr
 &= \f{i^{-\ell}}{2\pi} (-)^{s+1} \sum_{n=0}^{s}  \sum_{\ell=0}^\infty  
  \left(\begin{matrix}
s+\ell-n \\
\ell
\end{matrix}\right)
 D^{s-n} \left[\mathscr{A}_+(\ell,z), D^{n}  \int_{-\infty}^\infty du  \f{u^{s+\ell} }{(s+\ell)!} F^*_{ +}(u,z) \right]_\sG
 \cr
&=-\f{i^{-s}}{2\pi} \sum_{\ell=0}^\infty \sum_{n=0}^{s} 
  \left(\begin{matrix}
s+\ell-n \\
\ell
\end{matrix}\right)
 D^{s-n} \left[\mathscr{A}_+(\ell,z), D^{n}  \mathscr{F}^*_+(s + \ell,z) \right]_\sG.
\ee
In the manipulations above  we used
\be
u^{s+\ell-n} (\p_u^{-1})^{n} 
=u^{s+\ell} (\hD +n-1)^{-1}_{n}
= (\hD +n-s-\ell-1)^{-1}_{n}u^{s+\ell}\,,
\ee
which follows from the list  \eqref{uDelta},
and the identity
\be
\f1{ (n-s-\ell-1)_{n}}=(-1)^{n} \f{(s+\ell-n)!}{(s+\ell)!}\,.
\ee

The cubic charge aspects can be expressed as
\be
 \ym^3_s(z)=  \ym^3_{s+}(z)+ \ym^{3*}_{s-}(z)\,,
\ee
with
\be
 \ym^3_{s+}(z)&=
-
 \f1{(2\pi)^2}
 \sum_{n=1}^{s}  \sum_{k=0}^{n-1} \sum_{\ell=0}^\infty \sum_{m=0}^\infty
 \frac{(-)^{s-n} i^{-\ell-m}}{(s-n)!\ell! m!}
 \cr
 &\times
 D^{s-n}  \left[ \mathscr{A}_+(\ell,z),   \int_{-\infty}^\infty\rd u \, u^{s+\ell-n}(\p_u^{-1})^{k+1} D^k  \left[ u^m \mathscr{A}_+  (m,z) ,(\p_u^{-1})^{n-k-1} D^{n-k-1}F^*_{ +}(u,z)\right]_\sG\right]_\sG
 \cr
 &=
-
 \f1{(2\pi)^2} \sum_{n=1}^{s}  \sum_{k=0}^{n-1} \sum_{\ell=0}^\infty \sum_{m=0}^\infty
 \frac{(-)^{s-n}i^{-\ell-m}}{(s-n)!\ell! m!}
 \cr
 &\times
 D^{s-n}  \left[ \mathscr{A}_+(\ell,z),     D^k  \left[ \mathscr{A}_+  (m,z) , \int_{-\infty}^\infty\rd u\, 
 \f{ u^{s+\ell+k+m-n+1} }{ (n-s-\ell-1)_{k+1}}(\p_u^{-1})^{n-k-1} D^{n-k-1} F^*_{ +}(u,z)\right]_\sG\right]_\sG
 \cr
 &=
-
 \f1{(2\pi)^2} \sum_{n=1}^{s}  \sum_{k=0}^{n-1} \sum_{\ell=0}^\infty \sum_{m=0}^\infty
(-)^{s}i^{-\ell-m}
  \f{(s+\ell-n)!}{(s-n)! \ell!}
\f{(s+\ell+k+m-n+1)!}{(s+\ell+k-n+1)!m!}
 \cr
 &\times
 D^{s-n}  \left[ \mathscr{A}_+(\ell,z),     D^k  \left[ \mathscr{A}_+ (m,z) ,
 D^{n-k-1} \int_{-\infty}^\infty\rd u\,  \f{u^{s+\ell+m}}{(s+\ell+m)!} F^*_{ +}(u,z)\right]_\sG\right]_\sG\cr
 &=
-
 \f{i^{-s}}{(2\pi)^2}
 \sum_{\ell=0}^\infty \sum_{m=0}^\infty  \sum_{n=1}^{s}  \sum_{k=0}^{n-1}
 \left(\begin{matrix}
s+\ell-n \\
\ell
\end{matrix}\right)
 \left(\begin{matrix}
s+\ell+k+m-n+1 \\
m
\end{matrix}\right)
 \cr
 &\times
  D^{s-n}  \left[ \mathscr{A}_+(\ell,z)  ,   D^k  \left[ \mathscr{A}_+  (m,z) ,
 D^{n-k-1} \mathscr{F}^*_+(s+\ell+m,z)\right]_\sG\right]_\sG\,,\la{r3app}
\ee
where we used
\be
&u^{s+\ell-n}(\p_u^{-1})^{k+1} 
=u^{s+\ell+k-n+1} (\hD +k)^{-1}_{k+1}
= (\hD +n-s-\ell-1)^{-1}_{k+1}u^{s+\ell+k-n+1}\,,
\\
&(n-s-\ell-1)^{-1}_{k+1} =(-)^{k+1} \f{(s+\ell-n)!}{(s+\ell-n+k+1)!}
\,,
\ee
and
\be
&u^{s+\ell+k+m-n+1} (\p_u^{-1})^{n-k-1}
=u^{s+\ell+m}(u^{-1})^{n-k-1} (\p_u^{-1})^{n-k-1}\cr
&=u^{s+\ell+m} (\hD +n-k-2)^{-1}_{n-k-1}
=(\hD +n-k-s-\ell-m-2)^{-1}_{n-k-1}u^{s+\ell+m}\,,\\
&(n-k-s-\ell-m-2)^{-1}_{n-k-1}
=(-)^{n+k+1}\f{(s+k+\ell+m-n+1)!}{(s+\ell+m)!}\,.
\ee
We now perform this series of manipulations: we replace $k\to k-1$, we then switch sums $\sum_{n=1}^{s}  \sum_{k=1}^{n}= \sum_{k=1}^{s}  \sum_{n=k}^{s}$, and we  replace $n\to s-n$ to obtain
\be
\ym^3_{s+}(z)&=
-
\f{i^{-s}}{(2\pi)^2}
 \sum_{\ell=0}^\infty \sum_{m=0}^\infty  \sum_{k=1}^{s} \sum_{n=0}^{s-k}  
 \left(\begin{matrix}
\ell+n \\
\ell
\end{matrix}\right)
 \left(\begin{matrix}
n+m+ \ell +k \\
m
\end{matrix}\right)
 \cr
 &\times
  D^{n}  \left[ \mathscr{A}_\pm(\ell,z)  ,   D^{k-1}  \left[ \mathscr{A}_\pm  (m,z) ,
 D^{s-n-k} \mathscr{F}^*_\pm(s+\ell+m,z)\right]_{\sG}\right]_{\sG}.
\ee
At this point we switch sums again $\sum_{k=1}^{s} \sum_{n=0}^{s-k}=\sum_{n=0}^{s-1} \sum_{k=1}^{s-n}  $ and perform the final replacement $k \rightarrow k+1-n$  to arrive at
\be
\ym^3_{s\pm}(z)
 &=
-
 \f{i^{-s}}{(2\pi)^2}
 \sum_{\ell=0}^\infty \sum_{m=0}^\infty  \sum_{n=0}^{s-1}  \sum_{k=n}^{s-1} 
 \left(\begin{matrix}
\ell+n \\
\ell
\end{matrix}\right)
 \left(\begin{matrix}
m+ \ell +k+1 \\
m
\end{matrix}\right)
 \cr
 &\times
  D^{n}  \left[ \mathscr{A}_\pm(\ell,z)  ,   D^{k-n}  \left[ \mathscr{A}_\pm  (m,z) ,
 D^{s-k-1} \mathscr{F}^*_\pm(s+\ell+m,z)\right]_{\sG}\right]_{\sG}.
 \ee

\subsection{Corner charges}\la{App:YMcoch}

 We start proving the useful identity
 \be 
\int_{S} {\rm Tr} \left( 
 [P_s(\ell+1; \tau),  B(z)]_{\sG} \,A(z) \right)=
 (-)^{s+1} 
 \int_{S}  
 {\rm Tr}\left(   
 \left[P_s(-(s+\ell+1); \tau), A(z)\right]_{\sG}   B(z) 
\right),
\la{trans-app}
\ee 
where
\be
P^a_s(\alpha; \tau):=  \sum_{n=0}^{s}  \f{(-\alpha)_n}{n!} (D^{n}\tau^a_s(z))  D^{s-n}.
\ee
Starting from the LHS of \eqref{trans-app} and integrating by parts, we have
\be
&\int_{S} {\rm Tr} \left( 
 [P_s(\ell+1; \tau),  B(z)]_{\sG} \,A(z) \right)
 \cr
 &=
 \sum_{n=0}^{s} \sum_{m=0}^{s-n}(-)^{s+n}\f{(-(\ell+1))_n}{n!}
 \f{(s-n)_m}{m!}
 \int_{S} {\rm Tr} \left( 
 [  D^{s-m}\tau(z) ,    B(z)]_{\sG} \,D^m A(z) \right)
 \cr
 &=
\sum_{m=0}^{s} \sum_{n=0}^{s-m} (-)^{s}\f{(\ell+n)_n}{n!}
 \f{(s-n)_m}{m!}
 \int_{S} {\rm Tr} \left( 
 [  D^{s-m}\tau(z) ,    B(z)]_{\sG} \,D^m A(z) \right).
\ee
We now use
\be
 \sum_{n=0}^{s-m}\f{(\ell+n)_n}{n!}
 \f{(s-n)_m}{m!}
 =\f{(s+\ell+1)_{s-m}}{(s-m)!},
\ee
to write
\be
&\int_{S} {\rm Tr} \left( 
 [P_s(\ell+1; \tau),  B(z)]_{\sG} \,A(z) \right)
 \cr
& =
(-)^s \sum_{m=0}^{s}
\f{(s+\ell+1)_{s-m}}{(s-m)!}
 \int_{S} {\rm Tr} \left( 
 [  D^{s-m}\tau(z) ,    B(z)]_{\sG} \,D^m A(z) \right)
 \cr
& =
- (-)^s \sum_{m=0}^{s}
\f{(s+\ell+1)_{m}}{m!}
 \int_{S} {\rm Tr} \left( 
 [  D^{m}\tau(z) ,  D^{s-m} A(z)  ]_{\sG} \, B(z) \right)
  \cr
& =
  (-)^{s+1} 
 \int_{S}  
 {\rm Tr}\left(   
 \left[P_s(-(s+\ell+1); \tau), A(z)\right]_{\sG}   B(z) 
\right),
\ee
where in the second equality we used the cyclicity of the trace.

In terms of the operator \eqref{Ps}, the quadratic charges can be written as
\be
\YR^{2}_{s+}(\tau)&= - \f{(-i)^{s}}{2\pi}\sum_{\ell=0}^\infty \sum_{n=0}^{s} 
(-)^{n}
  \left(\begin{matrix}
\ell+n \\
\ell
\end{matrix}\right)
\int_{S} \tr\left(D^{n}\tau(z) 
 \left[ \mathscr{A}_+(\ell,z), D^{s-n}\mathscr{F}^{\dagger }_+(s+\ell,z)\right]_{\sG}\right) \cr
&= -  \f{i^{-s}}{2\pi} \sum_{\ell=0}^\infty \sum_{n=0}^{s} 
\f{(-(\ell+1))_n}{n!}
\int_{S} \tr\left(D^{n}\tau(z) 
 \left[ \mathscr{A}_+(\ell,z), D^{s-n}\mathscr{F}^{\dagger }_+(s+\ell,z)\right]_{\sG}\right) \cr
&= \f{i^{-s}}{2\pi} \sum_{\ell=0}^\infty  
\int_{S} \tr\left(\mathscr{A}_+(\ell,z) 
  \left[P_s(\ell+1; \tau) , \mathscr{F}^{\dagger}_+(s+\ell,z)\right]_\sG\right) \,,
\ee
where we used
\be\la{bella}
(-)^n
  \left(\begin{matrix}
\ell+n \\
\ell
\end{matrix}\right)
=\f{(-(\ell+1))_n}{n!}\,.
\ee

Using the identity \eqref{trans-app}, this can also be written as
\be
\YR^{2}_{s+}(\tau)
&=-  \f{i^{s}}{2\pi}\sum_{\ell=0}^\infty 
\sum_{k=0}^{s} 
\f{(\ell+s+1)_{k}}{ k!}
\int_{S} \tr\left(D^{k}\tau(z) 
 \left[D^{s-k}\mathscr{A}_+(\ell,z),  \mathscr{F}^{\dagger }_+(s+\ell,z) \right]_{\sG}\right)
 \cr
 &= -\f{i^{s}}{2\pi}   \sum_{\ell=s}^\infty 
\int_{S}  
 {\rm Tr}\left(   
 \left[P_s(-(\ell+1); \tau), \mathscr{A}_\pm(\ell-s,z)\right]_{\sG}   \mathscr{F}^\dagger_\pm(\ell,z) 
\right).
\ee

The quadratic charge action on $ \mathscr{F}^{\dagger b}_+(n,z)$ can then be computed from the commutator \eqref{com} as
\be
[\YR^{2}_{s+}(\tau),  \mathscr{F}^{\dagger b}_+(n,z)] &=
  \f{i^{-s}}{2\pi} \sum_{\ell=0}^\infty  
\int_{S}\rd^2z' \sqrt{\gamma}\, [ \mathscr{A}^a_+(\ell,z') , \mathscr{F}^{\dagger b}_+(n,z)]
 \left[P_s(\ell+1; \tau) ,\mathscr{F}^{\dagger }_+(s+\ell,z')\right]_{\sG a}
 \cr
 &=g_{\va \rm{YM}}^2  i^{-s}
 \left[P_s(n+1; \tau(z)) ,\mathscr{F}^{\dagger }_+(s+n,z)\right]_\sG^b\,. 
\ee

\section{YM algebra}\la{App:YMalg}

In this appendix we provide the proofs of the results presented in Section \ref{sec:YMqua-alg}. We make use of several identities involving the falling factorial
\be 
(x)_n = x(x-1)\cdots(x-n+1)= \frac{\Gamma( x+1)}{\Gamma(x-n+1)}.
\ee
First we use that 
\be
(-x-1)_n = (-1)^n (x+n)_n .
\ee 
We make use of  the fundamental binomial identity for the falling factorial
\be
\sum_{n=0}^{\infty} \frac{(x)_n}{n!} \frac{(y)_{s-n}}{(s-n)!} = \frac{(x+y)_s}{s!},
\ee 
where $s!:=\Gamma(s+1)$.
This identity can be proven  by recurrence from the shift identity satisfied by the falling factorial 
\be
\Delta (x)_n=n(x)_{n-1}, \qquad \Delta f(x):= f(x+1)-f(x)
\ee
and  the normalisation conditions  $(0)_n=\delta_{n 0}$.
This identity is valid for $-s\in \C\backslash \N$ and it can be shown to be equivalent to the Gauss hypergeometric identity.
When  $s\in \N$ it becomes more simply
\be 
\label{eq:binomial}
\sum_{n=0}^{s} \frac{(x)_n}{n!} \frac{(y)_{s-n}}{(s-n)!} = \frac{(x+y)_s}{s!}.
\ee
Other useful identities when $s$ is an integer can be obtained by taking $y$ to be a positive or negative integer. Such limits imply, after using  $\frac{(-(b+1))_n}{n!}= (-1)^n \left(\begin{matrix}
b+n \\
b
\end{matrix}\right)$, that 
\be \label{sumid2}
\frac{(x+b)_s}{s!}=\sum_{n=0}^s
\left(\begin{matrix}
b \\
n 
\end{matrix}\right) \frac{(x)_{s-n}}{(s-n)!}.\qquad 
\frac{(x-b-1)_s}{s!}=\sum_{n=0}^s (-)^n
 \left(\begin{matrix}
 b+n \\
 b 
 \end{matrix}\right) 
 \frac{(x)_{s-n}}{(s-n)!}, 
\ee 
where $b\in \mathbb{N}$.
Taking  $x$ to be a positive  integer and interchanging $n\to s-n$ gives the identities 
\be \label{sumid3}
 \left(\begin{matrix}
a+b  \\
s 
\end{matrix}\right)=\sum_{n=0}^s
\left(\begin{matrix}
a \\
n 
\end{matrix}\right)
\left(\begin{matrix}
b \\
s-n 
\end{matrix}\right) ,\qquad 
\frac{(a-b-1)_s}{s!}=\sum_{n=0}^s (-)^n
\left(\begin{matrix}
a \\
n 
\end{matrix}\right)
\left(\begin{matrix}
b+s-n \\
b 
\end{matrix}\right) ,
\ee 
where $a,b\in \mathbb{N}$.
Finally taking  $x$ to be a negative  integer gives the identity 
\be \label{sumid4}
\left(\begin{matrix}
a+b +s \\
s 
\end{matrix}\right)
&=\sum_{n=0}^s 
\left(\begin{matrix}
a+n-1 \\
a-1 
\end{matrix}\right)
\left(\begin{matrix}
b+s-n \\
b 
\end{matrix}\right) 
 .
\ee

\subsection{Quadratic charge commutator}
\la{App:YMalg-qua}

In this appendix we present a detailed computation of $I_1$ in Section \ref{sec:quad}. 
After changing variables $p \rightarrow p - m - n$ \eqref{I1} becomes
\be 
\label{eq:intermediate}
\begin{split}
I_1 &= \frac{i^{-s-s'}}{2\pi} g_{YM}^2\sum_{\ell = 0}^{\infty} \sum_{n = 0}^{s'} \sum_{m = 0}^s \sum_{p = m+n}^{s' +m} \frac{(-\ell - 1)_n}{n!} \frac{(-s' -1 -\ell)_m}{m!} \left(\begin{matrix} s' - n\\
p - n - m 
\end{matrix} \right) \\
&\times \int_{S}\Tr\left( [D^n \tau',[ D^{p-n} \tau, D^{s+s' -p}  \mathscr{F}^{\dagger}_{+}(s' + s+ \ell)]_{\sG}]_{\sG} \mathscr{A}_+(\ell)\right).
\end{split}
\ee
In this parametrization we see that the integral factor does not depend on $m$. 
The goal is therefore to perform the sum over $m$. 
We can do that after interchanging sums 
\be
\sum_{m = 0}^s \sum_{p = m + n}^{s' + m} = \sum_{p = n}^{s + s'} \sum_{m ={\rm max}[ 0,p -s']}^{{\rm min}[p - n, s]} .
\ee
The sum involves the binomial coefficient $\left(\begin{matrix} s' - n\\
p - n - m 
\end{matrix} \right)$ which vanishes when $ m< p-s'$. This means that we can replace the lower bound $m ={\rm max}[ 0,p -s']$ simply by $m=0$.
Therefore we  have two cases to evaluate.

{\bf Case I:} When  $p - n  \leq s$  we find  
\be 
\begin{split}
 &\sum_{m = 0}^{p - n} \frac{(-s' -1 -\ell)_m}{m!} \left(\begin{matrix} s' - n\\
p - n - m
\end{matrix} \right) = 
 \frac{(-1-\ell - n)_{ p-n}}{(p-n)!}
= (-1)^{p-n}\left(\begin{matrix} \ell+p\\
p-n
\end{matrix} \right),
\end{split}
\ee
where we have used \eqref{eq:binomial} with $x\to -(s'+\ell+1)$, $y\to (s'-n)$ and $s \to (p-n)$.\\

{\bf Case II:} for $p - n > s$ we find
\be \label{case2}
\begin{split}
& \sum_{m = 0}^{s}  \frac{(-s' -1 -\ell)_m}{m!} \left(\begin{matrix} s' - n\\
p - n - m
\end{matrix} \right) =  (-1)^{p+n} \left(\begin{matrix}
\ell + p\\
p - n
\end{matrix} \right)-
\sum_{m = s+1}^{p-n}  \frac{(-s' -1 -\ell)_m}{m!} \left(\begin{matrix} s' - n\\
p - n - m
\end{matrix} \right)
\\
=&  (-1)^{p+n} \left(\begin{matrix}
\ell + p\\
p - n
\end{matrix} \right)
+ (-1)^s \left(\begin{matrix} s' - n\\
p - s - 1 - n
\end{matrix} \right) 
\left(\begin{matrix}
\ell + s'+s+1\\
s+1
\end{matrix} \right)
\end{split}
\ee
where we defined 
\be 
{F}_1(s, s') :=  {}_3{F}_2\left[1, 1 + n - p + s, 2 + \ell + s  + s'; s + 2, 2 - p + s + s';1 \right]
\ee
and ${}_3{F}_2$ is the generalized hypergeometric function.\\

We conclude that the final result of \eqref{eq:intermediate} includes two contributions. We will denote the common contribution to the two cases, namely the one proportional to $\left(\begin{matrix}
\ell + p\\
p - n
\end{matrix} \right)$ by $J_1$. We then find that $I_1= J_1+ J'_1$
where 
\be 
J_1 =&  \frac{i^{-s-s'}g_{YM}^2}{2\pi} \sum_{\ell = 0}^{\infty} \sum_{n=0}^{s'}  \sum_{p = n}^{s+s'} (-1)^p \left(\begin{matrix} p\\
n
\end{matrix} \right) \left(\begin{matrix} \ell + p\\
\ell
\end{matrix} \right)\cr
&\times\int_{S}\Tr\left( [D^n \tau',[ D^{p-n} \tau, D^{s+s' -p}  \mathscr{F}^{\dagger}_{+}(s' + s+ \ell)]_{\sG}]_{\sG} \mathscr{A}_+(\ell)\right),
\cr
J_1'=& \frac{i^{-s-s'}}{2\pi} g_{YM}^2\sum_{\ell = 0}^{\infty} \sum_{n = 0}^{s'-1} \sum_{p = n+s+1}^{s' + s}  (-1)^{s+n} \left(\begin{matrix} \ell + n\\
n \end{matrix} \right)\left(\begin{matrix} 1 + \ell + s + s'\\
s + 1 \end{matrix} \right)\left(\begin{matrix} s'- n\\
p - n - s - 1\end{matrix} \right) \cr
&\times  {F}_1(s,s') \int_{S} \Tr\left([D^n \tau', [ D^{p-n} \tau, D^{s+s' -p}  \mathscr{F}^{\dagger}_{+}(s' + s+ \ell) ]_{\sG}]_{\sG} \mathscr{A}_+(\ell)\right) .
\ee 
Note that the sum over $n$ in $J_1'$ only goes up to $s' - 1$ due to the fact that $p \geq n + s + 1$.
In the first equality we used that
\be 
 \frac{(\ell + n)_n}{n!} \left(\begin{matrix}
   \ell + p\\
   p - n
\end{matrix} \right) = \left(\begin{matrix}
   \ell + p\\
   \ell
\end{matrix} \right)\left(\begin{matrix}
   p\\
    n
\end{matrix} \right).
\ee

\subsection{First term in cubic algebra}
\label{app:cubic}

In this appendix we evaluate $I_3$.
We start with \eqref{eq:I3} (with $s \leftrightarrow s' $ and $\tau \leftrightarrow \tau'$) and shift
 \be 
 k \rightarrow k - s' - 1
 \ee
 to find 
 \be 
 \begin{split}
 I_3(s', \tau'; s, \tau) &= \frac{i^{-s+s'}}{2\pi} g_{YM}^2  \sum_{m = 0}^{\infty} \sum_{n = 0}^{s-1} \sum_{k = n+ s' + 1}^{s + s'} \sum_{p = 0}^{k - n - s' - 1} (-1)^{k+s' + 1} \left(\begin{matrix}
 s' + n\\
 s'
 \end{matrix} \right)\left(\begin{matrix}
 m+ k\\
 m
 \end{matrix} \right) \\
 &\times \left(\begin{matrix}
 k - n -s' -1\\
 p
 \end{matrix} \right)\int_S \Tr\left( [D^{n + p} \tau, D^{k - n - p} \tau']_{\sG} [\mathscr{A}(m), D^{s+s' - k } \mathscr{F}^{\dagger }(s + s' + m)]_{\sG} \right).
 \end{split}
 \ee
 We exchange  sums
 \be 
 \sum_{n = 0}^{s-1} \sum_{k = n+ s' + 1}^{s + s'} \sum_{p = 0}^{k - n - s' - 1} = \sum_{k = s' + 1}^{s + s' } \sum_{n = 0}^{k - s' - 1} \sum_{p = 0}^{k - n - s' - 1} = \sum_{k = s' + 1}^{s + s'}  \sum_{p = 0}^{k - s' - 1} \sum_{n = 0}^{ k - s' - 1-p} ,
 \ee
 where in the first exchange we used that $k-s'-1 < s$  so there is no need for another bound.
Shifting $n \rightarrow n - p$,  and changing the order of the commutator, we find
 \be 
 \begin{split}
 I_3(s', \tau'; s, \tau) &= \frac{i^{-s+s'}}{2\pi} g_{YM}^2  \sum_{m = 0}^{\infty}  \sum_{k = s' + 1}^{s + s'}  \sum_{p = 0}^{k - s' - 1} \sum_{n = p}^{k - s' - 1}  (-1)^{k+s'} \left(\begin{matrix}
 s' + n - p\\
 s'
 \end{matrix} \right)\left(\begin{matrix}
 m+ k\\
 m
 \end{matrix} \right) \\
 &\times \left(\begin{matrix}
 k - n -s' -1 + p\\
 p
 \end{matrix} \right)  \int_S \Tr\left( [D^{n } \tau, D^{k - n } \tau']_{\sG} [ D^{s+s' - k } \mathscr{F}^{\dagger }(s + s' + m), \mathscr{A}(m)]_{\sG} \right).
 \end{split}
 \ee
 Now we switch sums again 
 \be 
 \sum_{p = 0}^{k - s' - 1} \sum_{n = p}^{k-s'-1} = \sum_{n = 0}^{k - s'-1} \sum_{p = 0}^n
 \ee
 and do the sum on $p$ using \eqref{sumid4} with $a\to k-s'-n$, $b\to s'$ and $s\to n$,
 \be 
 \sum_{p = 0}^n\left(\begin{matrix}
 s' + n - p\\
 s'
 \end{matrix} \right)\left(\begin{matrix}
 k  -s' + p-n-1\\
 p
 \end{matrix} \right) = \left(\begin{matrix}
 k\\
 n
 \end{matrix} \right) .
 \ee
 We are then left with a double sum 
 \be 
 \sum_{k = s' + 1}^{s + s'} \sum_{n = 0}^{k - s'-1}=
  \sum_{n = 0}^{s-1} \sum_{k = n+ s' + 1}^{s + s'}.
 \ee 
 The result \eqref{I3-final} follows immediately  after exchanging $(s,\tau)$ with $(s',\tau')$.

 \subsection{Second term in cubic algebra}
\label{app:cubic2}

We next show that $I_4$ simplifies to \eqref{simp4}. Starting with \eqref{I4-in} (with $(s, \tau) \rightarrow (s', \tau')$) and letting $p \rightarrow p - n - s' - 1$, we find
 \be 
 \begin{split}
 I_4{(s', \tau'; s, \tau)} &= \frac{i^{s' - s}}{2\pi} g_{YM}^2  \sum_{\ell = 0}^{\infty} \sum_{n = 0}^{s-1} \sum_{k = n}^{s - 1}
 \sum_{p = n + s' + 1}^{k + s' + 1}(-1)^n \left(\begin{matrix} 
 \ell + n\\
 \ell
 \end{matrix}\right) \left(\begin{matrix} 
 s' + k + \ell +1\\
 s'
 \end{matrix}\right) \left(\begin{matrix} 
 k - n\\
 p - n - s' - 1
 \end{matrix} \right) \\
 &   \int_S\Tr\left( [D^n\tau, \mathscr{A}_+(\ell)]_{\sG}  [D^{p - n} \tau', D^{s + s' - p} \mathscr{F}^{\dagger}_+(s + \ell + s')]_{\sG} \right).
 \end{split}
 \ee
 We can now switch sums
 \be 
 \sum_{k = n}^{s - 1} \sum_{p = n + s' + 1}^{k + s' + 1} = \sum_{p = n + s' + 1}^{s + s'} \sum_{k = p - s' - 1}^{s - 1}
 \ee 
 and evaluate the sum on $k$
\be
\label{sum-explicit}
 \begin{split}
& \sum_{k = p - s' - 1}^{s - 1} \left(\begin{matrix} 
 s' + k + \ell +1\\
 s'
 \end{matrix}\right) \left(\begin{matrix} 
k - n\\
 p - n - s' - 1
 \end{matrix}\right) \\
 &=  \sum_{k = p - s' - 1}^{s - 1} (-1)^{1+k+p+s'} \left(\begin{matrix} 
 s' + k + \ell +1\\
 s'
 \end{matrix}\right) \left(\begin{matrix} 
n+s'-p\\
n-k-1
 \end{matrix}\right)\\
 &= \left(\sum_{k = p - s' - 1}^{n - 1} + \sum_{k = n}^{s-1} \right) (-1)^{1+k+p+s'} \left(\begin{matrix} 
 s' + k + \ell +1\\
 s'
 \end{matrix}\right) \left(\begin{matrix} 
n+s'-p\\
n-k-1
 \end{matrix}\right)\\
 &= \sum_{k = p - s' + \ell}^{n + \ell} (-1)^{k+p+s' + \ell} \left(\begin{matrix} 
 s' + k \\
 s'
 \end{matrix}\right) \left(\begin{matrix} 
n+s'-p\\
n-k + \ell
 \end{matrix}\right) + \sum_{k = n+\ell+1}^{s+ \ell} (-1)^{\ell + k+p+s'} \left(\begin{matrix} 
 s' + k \\
 s'
 \end{matrix}\right) \left(\begin{matrix} 
n+s'-p\\
n-k + \ell
 \end{matrix}\right),\\
 \end{split}
\ee
where in the last equality we shifted $k \rightarrow k - \ell - 1$. We now use the second binomial identity in \eqref{sumid2} which can be written as
\be 
\sum_{n = 0}^s (-1)^n  \left(\begin{matrix} 
 b +n\\
 b
 \end{matrix}\right) \left(\begin{matrix} 
x\\
s - n
 \end{matrix}\right) = \frac{(x - b - 1)_s}{s!} = \left(\begin{matrix} 
 x-b-1\\
 s
 \end{matrix}\right)
\ee
to simplify the first sum in \eqref{sum-explicit}
\be 
\begin{split}
&\sum_{k = p - s' + \ell}^{n + \ell} (-1)^{k+p+s' + \ell} \left(\begin{matrix} 
 s' + k \\
 s'
 \end{matrix}\right) \left(\begin{matrix} 
n+s'-p\\
n-k + \ell
 \end{matrix}\right) \\
 &= \left(\sum_{k=0}^{n + \ell} - \sum_{k = 0}^{p - s' + \ell - 1} \right) (-1)^{k+p+s' + \ell} \left(\begin{matrix} 
 s' + k \\
 s'
 \end{matrix}\right) \left(\begin{matrix} 
n+s'-p\\
n-k + \ell
 \end{matrix}\right) \\
 &= (-1)^{p+s'+\ell}\left(\begin{matrix} 
n -p - 1\\
n+\ell
 \end{matrix}\right) - \sum_{k = 0}^{p - s' + \ell - 1} (-1)^{k+p+s' + \ell} \left(\begin{matrix} 
 s' + k \\
 s'
 \end{matrix}\right) \left(\begin{matrix} 
n+s'-p\\
n-k + \ell
 \end{matrix}\right)\\
 &= (-1)^{p + s' + n} \left(\begin{matrix} 
p+\ell\\
p - n
 \end{matrix}\right) - \sum_{k = 0}^{p - s' + \ell - 1} (-1)^{k+p+s' + \ell} \left(\begin{matrix} 
 s' + k \\
 s'
 \end{matrix}\right) \left(\begin{matrix} 
n+s'-p\\
n-k + \ell
 \end{matrix}\right).
\end{split}
\ee 
Putting everything together we find
\be 
&\sum_{k = p - s' - 1}^{s - 1} \left(\begin{matrix} 
 s' + k + \ell +1\\
 s'
 \end{matrix}\right) \left(\begin{matrix} 
k - n\\
 p - n - s' - 1
 \end{matrix}\right) = (-1)^{n+p+s'} \left(\begin{matrix} 
 \ell+p\\
 p-n
 \end{matrix}\right)\\
 &+ \left(\sum_{k = n+\ell+1}^{s+ \ell} - \sum_{k = 0}^{p - s' + \ell - 1} \right) (-1)^{\ell + k+p+s'} \left(\begin{matrix} 
 s' + k \\
 s'
 \end{matrix}\right) \left(\begin{matrix} 
n+s'-p\\
n-k + \ell
 \end{matrix}\right)
\ee
or equivalently
 \be
 \begin{split}
& \sum_{k = p - s' - 1}^{s - 1} \left(\begin{matrix} 
 s' + k + \ell +1\\
 s'
 \end{matrix}\right) \left(\begin{matrix} 
k - n\\
 p - n - s' - 1
 \end{matrix}\right) \\
 =& (-1)^{n+p+s'} \left(\begin{matrix} 
 \ell+p\\
 p-n
 \end{matrix}\right) -  \left(\begin{matrix} 
 s - n\\
 p - s' - 1 - n
 \end{matrix}\right)  \left(\begin{matrix} 
 s' +s + \ell +1\\
 s'
 \end{matrix}\right) {F_2(s', s)}\label{sumidF}
 \end{split}
 \ee
 where  we defined
 \be 
 F_2(s', s) \equiv {}_3F_2[1, 1 - n + s, 2 + l + s + s', 2 + \ell + s, 
   2 - p + s + s', 1].
 \ee
 This  equality which is similar to \eqref{case2} can also be checked by direct evaluation of the sum in \eqref{sum-explicit} with mathematica.
 
 We recover \eqref{simp41} from the first term in \eqref{sumidF}, while for \eqref{simp42} we use the second term in \eqref{sumidF}. In total  we obtain \eqref{simp4}. In both cases we acquire a sign upon cycling the terms in the trace and reordering a commutator.

\subsection{Putting everything together}
\label{app:everything}

We first prove an important identity relating hypergeometric functions. This will allow us to show that the hypergeometric functions appearing in $I_1$ and $I_4$ cancel. The relevant hypergeometric identities are
\be 
\label{hg-id}
\begin{split}
{}_3F_2[a_1,a_2,a_3;b_1,b_2; 1] &= \frac{\Gamma(b_1)\Gamma(b_1 + b_2 - a_1- a_2 - a_3)}{\Gamma(b_1 - a_1)\Gamma(b_1 + b_2 - a_2 -a_3)}\\
&\times{}_3F_2[a_1,b_2 - a_2, b_2 - a_3;b_1 + b_2 - a_2 - a_3,b_2; 1],\\
&\quad Re(b_1 + b_2 - a_1 -a_2 - a_3) > 0, \quad Re(b_1 - a_1) > 0, \\
{}_3F_2[-n,b,c;d,e; 1] &= \frac{(d-b)^{(n)}}{(d)^{(n)}}{}_3F_2[-n,b,e-c;e,b-d-n+1; 1], \quad n \in \mathbb{Z}_+,
\end{split}
\ee
   where $(\alpha)^{(n)}=\Gamma(\alpha+n)/\Gamma(\alpha)$ is the raising factorial.
The second of these can be used to rewrite
\be 
F_1(s,s') = \frac{(s + 2 - 1)^{(p - 1 - n - s)}}{(s+2)^{(p - 1 - n - s)}} {}_3 F_2[1,1 - p +s+n,-p-\ell;2-p+s+s',1+n-p;1].
\ee
We could apply it because $1 + n - p + s < 0$ for the relevant summation range. 
We now use the first identity in \eqref{hg-id} to rewrite
\be 
\begin{split}
F_1(s,s') &= \frac{(s + 2 - 1)^{(p - 1 - n - s)}}{(s+2)^{(p - 1 - n - s)}} \frac{(n - p)}{(1 + s' + \ell)} \\
&\times {}_3 F_2[1,1 + s' - n, 2 + s + s' + \ell;2-p+s+s',2 + s' + \ell;1] \\
&= \frac{(s + 1)}{(p - n)} \frac{(n - p)}{(1+ s' + \ell)} F_2(s,s') = -\frac{(s + 1)}{(1+ s' + \ell)} F_2(s,s'),
\end{split}
\ee
which is exactly what we need for these terms to cancel in the sum.

We next outline the steps leading from \eqref{R22} to \eqref{eq:final-YM}.
We start by shifting $p \rightarrow p + n$ in the first line of \eqref{R22} which expresses $J_1+J_4$ and antisymmetrizing to get
\be 
\begin{split}
    &\sum_{n = 0}^{s'} \sum_{p = 0}^s (-1)^{p+n}\left(\begin{matrix}
        \ell + p + n\\
        \ell
    \end{matrix} \right)\left(\begin{matrix}
      p + n\\
        n
    \end{matrix} \right) \\
    &\times \int_S \Tr \left([D^n\tau',[D^p\tau, D^{s+s'-n-p}\mathscr{F}_+^{\dagger}(s+s'+\ell)]_\sG]_\sG\mathscr{A}_+(\ell)\right)  \\
    &- \Tr\left([D^n\tau,[D^p\tau', D^{s+s'-n-p}\mathscr{F}_+^{\dagger}(s+s'+\ell)]_\sG]_\sG \mathscr{A}_+(\ell)\right)\\
    &= \sum_{n = 0}^{s'} \sum_{p = 0}^s (-1)^{p+n}\left(\begin{matrix}
        \ell + p + n\\
        \ell
    \end{matrix} \right)\left(\begin{matrix}
      p + n\\
        n
    \end{matrix} \right) \int_S \Tr \left([[D^n\tau', D^p \tau]_\sG,D^{s+s'-n-p}\mathscr{F}_+^{\dagger}(s+s'+\ell)]_\sG\mathscr{A}_+(\ell)\right) \\
      &= \sum_{n = 0}^{s'} \sum_{p = n}^{s+n} (-1)^{p}\left(\begin{matrix}
        \ell + p \\
        \ell
    \end{matrix} \right)\left(\begin{matrix}
      p \\
        n
    \end{matrix} \right)  \int_S \Tr \left([[D^n\tau', D^{p-n} \tau]_\sG,D^{s+s'-p}\mathscr{F}_+^{\dagger}(s+s'+\ell)]_\sG\mathscr{A}_+(\ell)\right).
\end{split}
\ee
As a result, we see that \eqref{R22}  simplifies to
\be 
\begin{split}
&[R_{s+}(\tau), R_{s'+}(\tau')]^2 \\
&=  \frac{i^{-s-s'}}{2\pi} g_{YM}^2\sum_{\ell = 0}^{\infty}
\left(  \sum_{n = 0}^{s'}  \sum_{p = n}^{s+n} +   \sum_{p = s + 1}^{s + s'}  \sum_{n = 0}^{p -s-1} +  \sum_{p = s' + 1}^{s + s'} \sum_{n = 1+s'}^{p}  \right)(-1)^p \left(\begin{matrix} p\\
n
\end{matrix} \right) \left(\begin{matrix} \ell + p\\
\ell
\end{matrix} \right) \\
&\times \int_{S} \Tr\left[ [D^n \tau', D^{p-n} \tau]_\sG [D^{s+s' -p}  \mathscr{F}^{\dagger}_{+}(s' + s+ \ell), \mathscr{A}_+(\ell)]_\sG \right].
\end{split}
\ee
The last two terms come from $I_3$ \eqref{I3-final} and its antisymmetrization upon using the following identity
\be 
\label{id-useful}
\begin{split}
&\sum_{p = s' + 1}^{s + s'} \sum_{n = 0}^{p - s' - 1} (-1)^p \left(\begin{matrix}
\ell + p\\
\ell
\end{matrix} \right)\left(\begin{matrix}
p\\
n
\end{matrix} \right) \int D^n \tau^a D^{p - n} \tau'^b D^{s + s' - p} \\
&= \sum_{p = s' + 1}^{s + s'} \sum_{n = 1+s'}^{p} (-1)^p \left(\begin{matrix}
\ell + p\\
\ell
\end{matrix} \right)\left(\begin{matrix}
p\\
n
\end{matrix} \right) \int D^{n - p} \tau^a D^{n} \tau'^b D^{s + s' - p} ,
\end{split}
\ee
which is easy to prove by making the change of variables variables $ n \to n - p$. 

Finally switching sums for the last term in \eqref{R22}, we notice that the sums can be rearranged as 
\be 
&\sum_{n = 0}^{s'}  \sum_{p = n}^{s+n} +   \sum_{p = s + 1}^{s + s'}  \sum_{n = 0}^{p -s-1} +  \sum_{p = s' + 1}^{s + s'} \sum_{n = 1+s'}^{p} = \sum_{n = 0}^{s'}  \sum_{p = n}^{s+n} +  \sum_{n = 0}^{s'-1} \sum_{p = n+ s + 1}^{s + s'}   +  \sum_{n = 1+s'}^{s+s'} \sum_{p = n}^{s + s'} \\
&= \sum_{n = 0}^{s + s'} \sum_{p = n}^{s + s'} - \sum_{n = 0}^{s' - 1} \sum_{p = s + n + 1}^{s+ s'} + \sum_{n = 0}^{s'-1} \sum_{p = n+ s + 1}^{s + s'} = \sum_{p=0}^{s+s'}
  \sum_{n=0}^{p}
\ee
from which \eqref{eq:final-YM} follows immediately.

\section{Gravity cubic charges}\la{App:GRcharges}

We focus on the positive helicity charge (same result can be derived for the negative one).
We start  from the general relation \eqref{qren} to write the cubic charge as
\be
Q^{3}_{s}(\tau)
&=-\f12 \sum_{\ell=2}^{s}
 \frac{ (\ell+1)}{(s-\ell)!}  \int_S \rd^2 z \sqrt{q}\,  D^{s-\ell}\tau \int_{-\infty}^\infty \rd u\, u^{s-\ell}
C (u,z) \cQ^{2}_{\ell-2 }(u,z).
\ee
We now use \eqref{uexp} again together with
\be 
\mathcal{Q}_s(u) = -\frac{1}{2}\left(i^{-s} \mathcal{Q}_{s+}(u) + i^s \mathcal{Q}^*_{s-}(u)\right)
\ee and the expression for the quadratic charge 
\be\la{Q2}
\cQ_{\ell-2+}^{2}(u,z)&= -i^s
{\frac{1}{2}}\sum_{m=0}^{\ell-2} (m+1)(\p_u^{-1})^{\ell-m-1}  D^{\ell-m-2} \left[C_+(u,z)  (\p_u^{-1})^{m-1}  D^{m} N_+^{*}(u,z)\right] \,,
\ee 
to write
\be
Q^{3}_{s+}(\tau)
&=\f{i^s}{{4}}
\sum_{\ell=2}^{s}\sum_{m=0}^{\ell-2}
 \frac{ (\ell+1)(m+1)}{(s-\ell)!}  \int_S \rd^2 z \sqrt{q}\,  D^{s-\ell}\tau
 \cr
 &\times
  \int_{-\infty}^\infty \rd u\, u^{s-\ell}
C_+ (u,z)(\p_u^{-1})^{\ell-m-1}  D^{\ell-m-2} \left[C_+(u,z)  (\p_u^{-1})^{m-1}  D^{m} N_+^{*}(u,z)\right] 
\cr
&=-\f{i^s}{{4}(2\pi)^2 }
\sum_{n=0}^\infty \sum_{k=0}^\infty
\sum_{\ell=2}^{s}\sum_{m=0}^{\ell-2}
i^{-n-k}
 \frac{ (\ell+1)(m+1)}{n!k!(s-\ell)!}  \int_S \rd^2 z \sqrt{q}\,  D^{s-\ell}\tau\, \mathscr{S}_+(n,z)
 \cr
 &\times
  \int_{-\infty}^\infty \rd u\,  u^{s+n-\ell}
(\p_u^{-1})^{\ell-m-1}  D^{\ell-m-2} \left[\mathscr{S}_+(k,z)  u^k (\p_u^{-1})^{m-1}  D^{m} N_+^{*}(u,z)\right].
\ee
We can now introduce the operator $\widehat\Delta :=\pa_u u$, which integrates to zero due to our  choice of boundary conditions, and use the property
\be
 (u^{-1})^{\ell-s-n} ( \p_u^{-1})^{\ell-m-1}   &= u^{s+n-m-1}  (u^{-1} \p_u^{-1})^{\ell-m-1} 
 =u^{s+n-m-1} (\hD+\ell-m-2)_{\ell-m-1}^{-1}\cr
 &=  (\hD+\ell-s-n-1)_{\ell-m-1}^{-1}u^{s+n-m-1}  \,,\cr
 (u^{-1})^{m+1 -s-n-k} ( \p_u^{-1})^{m-1}  &= u^{s+n+k-2}  (\hD+m-2)_{m-1}^{-1}
 = (\hD+m-s-n-k)_{m-1}^{-1} u^{s+n+k-2}, \cr
\ee
to write the cubic charges
\be
Q_{s+}^{3}(\tau)
&=-\f{i^s}{{4}(2\pi)^2 }\sum_{n=0}^{\infty} \sum_{k=0}^{\infty} \sum_{\ell=2}^s \sum_{m=0}^{\ell-2}
i^{-n-k}  \frac{ (\ell+1)(m+1)}{n!k!(s-\ell)!}  \int_S \rd^2 z \sqrt{q}\,  D^{s-\ell}\tau\, \mathscr{S}_+(n,z)
\cr
&\times 
 \int_{-\infty}^{\infty} \rd u 
   \f{u^{s+n-m-1} }{(\ell-s-n-1)_{\ell-m-1} }
  D^{\ell-m-2} \left[ \mathscr{S}_+(k,z)  u^k D^{m} ( \p_u^{-1})^{m-1} N_+^*(u,z) \right]
\cr
&=-\f{i^s}{{4}(2\pi)^2 } \sum_{n=0}^{\infty} \sum_{k=0}^{\infty} \sum_{\ell=2}^s \sum_{m=0}^{\ell-2}
i^{-n-k}  \frac{(\ell+1)  (m+1)}{(s-\ell)! n!k!}
\f{1}{(\ell-s-n-1)_{\ell-m-1}  (m-s-n-k)_{m-1}}\cr
&\times 
\int_S \rd^2 z \sqrt{q}\, D^{s-\ell} \tau(z) \mathscr{S}_+(n,z)
\left[
  D^{\ell-m-2} \left[   \mathscr{S}_+(k,z)D^{m}    \int_{-\infty}^{\infty} \rd u \,u^{s+n+k-2} N_+^*(u,z) \right]
  \right] \cr
&=\frac{(-)^{s}}{4 (2\pi)^2} \sum_{n=0}^{\infty} \sum_{k=0}^{\infty} \sum_{\ell=2}^s \sum_{m=0}^{\ell-2}(-)^{\ell} (\ell+1) (m+1)
  \left(\begin{matrix}
s+n-\ell\\
n
\end{matrix}\right)
 \left(\begin{matrix}
s+n+k-m-1\\
k
\end{matrix}\right)
\cr
&\times 
\int_S \rd^2 z \sqrt{q}\, D^{s-\ell} \tau(z)\mathscr{S}_+(n, z) 
\left[
  D^{\ell-m-2} \left[  \mathscr{S}_+(k, z) D^{m}  \mathscr{M}^*_+(s+n+k-2,z) \right]
  \right] \,,
\ee
where we have used \eqref{softC}
and the identity
\be
\f1{ (\alpha-n)_{\ell}}=(-1)^{\ell} \f{(n-\alpha-1)!}{(n-\alpha-1+\ell)!}\,.
\ee

We now use this expression to compute the action
\be
&[ Q^{3}_{s'+}(\tau'),  \mathscr{S}_+(n,z) ]= 
\frac{(-)^{s'}}{4 (2\pi)^2} \sum_{p=0}^{\infty} \sum_{k=0}^{\infty} \sum_{\ell=2}^{s'} \sum_{m=0}^{\ell-2} (-)^{\ell}(\ell+1) (m+1)
  \left(\begin{matrix}
s'+p-\ell\\
p
\end{matrix}\right)
 \left(\begin{matrix}
s'+p+k-m-1\\
k
\end{matrix}\right)
\cr
&\times 
\int_S \rd^2 z' \sqrt{q}\, D_{z'}^{s'-\ell} \tau'(z')\mathscr{S}_+(p, z') 
  D_{z'}^{\ell-m-2} \left[  \mathscr{S}_+(k, z') D_{z'}^{m} [ \mathscr{M}^\dagger_+(s'+p+k-2,z') ,  \mathscr{S}_+(n,z) ]\right]
 \cr
  &=-\frac{(-)^{s'}\k}{16\pi}  \sum_{k=0}^{\infty} \sum_{\ell=2}^{s'} \sum_{m=0}^{\ell-2} (-)^{\ell}  (\ell+1) (m+1)
  \left(\begin{matrix}
n-k+2-\ell\\
s'-\ell
\end{matrix}\right)
 \left(\begin{matrix}
n-m+1\\
k
\end{matrix}\right)
\cr
&\times 
\int_S \rd^2 z' \sqrt{q}\, D_{z'}^{s'-\ell} \tau'(z')\mathscr{S}_+(n-k-s'+2, z') 
  D_{z'}^{\ell-m-2} \left[  \mathscr{S}_+(k, z') D_{z'}^{m} \d^2(z,z') \right]
 \cr
  &=-\frac{(-)^{s'}\k}{16\pi}   \sum_{k=0}^{\infty} \sum_{\ell=2}^{s'} \sum_{m=0}^{\ell-2}\sum_{p=0}^{\ell-m-2}
  (\ell+1) (m+1)
  \left(\begin{matrix}
n-k+2-\ell\\
s'-\ell
\end{matrix}\right)
 \left(\begin{matrix}
n-m+1\\
k
\end{matrix}\right)
 \left(\begin{matrix}
\ell-m-2\\
p
\end{matrix}\right)
\cr
&\times 
D^{m}\left[
D^{s'-m-2-p} \tau'(z)  D^p\mathscr{S}_+(n-k-s' +2, z) 
\mathscr{S}_+(k, z)  
\right]
 \,.
\ee

\section{Gravity algebra}\la{App:GRalg}

\subsection{Linear order}\la{App:GRalg-lin}

We concentrate on the same helicity sector and compute the linear charge commutator
\be\la{QQ1}
[Q_{s}(\tau), Q_{s'}(\tau')]_+^1=[Q^{1}_{s+}(\tau), Q^{2} _{s'+}(\tau')]-[ Q^{1}_{s'+}(\tau'),Q^{2}_{s+}(\tau)]\,.
\ee
We use the Jacobi property
\be
&[  [Q^{1}_{s+}(\tau) ,Q^{2}_{s'+}(\tau') ],\mathscr{S}_+(n,z) ]
-[  [Q^{1}_{s'+}(\tau') ,Q^{2}_{s+}(\tau) ],\mathscr{S}_+(n,z) ]\cr\,
&=[Q^{1}_{s+}(\tau) , [ Q^{2}_{s'+}(\tau'),  \mathscr{S}_+(n,z) ]]-[Q^{1}_{s'+}(\tau') , [ Q^{2}_{s+}(\tau),  \mathscr{S}_+(n,z) ]],
\ee 
where we used the fact that, by means of \eqref{corn-comm} and \eqref{Q1}, \eqref{Q2+},   
\be
[Q^{2}_{s'+}(\tau') , [ Q^{1}_{s+}(\tau),  \mathscr{S}_+(n,z) ]] =0\,.
\ee
We thus start with the action \eqref{Q2Nn} and 
we recall the  identity \eqref{magic}
\be
 D^{\alpha-1} \left[(s+1) \tau D +(s+1 -  \alpha) D\tau\right] D^{s-\alpha}
 =\left[ \sum_{k=0}^s +\sum_{k=s+1}^\infty \right] (s+1-k) \frac{(\alpha)_{k}}{ k! }   (D^{k} \tau ) D^{s-k}\,,
\ee
that allows us to rewrite 
\be
[ Q^{2}_{s+}(\tau),  \mathscr{S}_+(n,z) ]&=
-\f{\kappa^2}{{4}}(-)^{s} 
 D^{n+2} \left[(s+1) \tau D +(s-n-2) D\tau\right] D^{s-n-3}
\mathscr{S}_+(n-s+1,z)
\cr
&+{ {R(s,n)} }
\,, \la{Q2Nn2}
\ee
where {$R(s,n)$} corresponds to the contribution in \eqref{magic} given by the sum $\sum_{k=s+1}^\infty$ which we compute shortly. 
It follows that
\be
&[Q^{1}_{s+}(\tau) , [ Q^{2}_{s'+}(\tau'),  \mathscr{S}_+(n,z) ]]=
-\frac{\kappa^2}{4}(-)^{s'}  
 D_z^{n+2} \left[(s'+1) \tau' D_z +(s'-n-2) D\tau'\right]\cr
 &\times
  \int_S  \rd^2 {z'} \sqrt{q}\, \tau(z') D_{z'}^{s+2} 
  D_z^{s'-n-3}
[ \mathscr{M}^\dagger_+(s,z'), \mathscr{S}_+(n-s'+1,z)]
+{[Q^1_{s+}(\tau), R(s',n)] }\cr
&=
\pi \frac{\kappa^4}{4} (-)^{s+s'}
 D_z^{s+s'+1} \left[(s'+1) \tau' D_z -(s+1) D\tau'\right] \tau(z)\d_{s+s',n+1}
 +{[Q^1_{s+}(\tau), R(s',n)] }.~~~~
 \, \la{Q2Nn-res}
\ee

Finally, we compute
\be
{[Q^1_{s+}(\tau), R(s',n)] }&=
\frac{\kappa^2}{4} (-)^{s'}  
 \sum_{k=s'+1}^{n+3}
 (s'-k+1 ) 
 \left(\begin{matrix}
n+3\\
k
\end{matrix}\right)
\cr
 &\times D_z^{k} \tau'(z)
  \int_S  \rd {z'}^2 \sqrt{q}\, \tau(z') D_{z'}^{s+2} 
  D_z^{s'-k}
[ \mathscr{M}^\dagger_\pm(s,z'), \mathscr{S}_+(n-s'+1,z)]\cr
&=
-\pi \frac{\kappa^4}{4}  (-)^{s+s'} 
 \sum_{k=s'+1}^{s+s'+2}
 (s'-k+1 ) 
 \left(\begin{matrix}
s+s'+2\\
k
\end{matrix}\right)
D^{k} \tau'(z)
 D^{s+s'+2-k}  \tau(z)\cr
 &=
\pi \frac{\kappa^4}{4} (-)^{s+s'}   
 \sum_{k=0}^{s}
 (s-k+1 ) 
 \left(\begin{matrix}
s+s'+2\\
k
\end{matrix}\right)
D^{k} \tau(z)
 D^{s+s'+2-k}  \tau'(z)\cr
 &=
[Q^{1}_{s'+}(\tau') , [ Q^{2}_{s+}(\tau),  \mathscr{S}_+(n,z) ]]\,.
\ee
Therefore we have
\be
&[Q^{1}_{s+}(\tau) , [ Q^{2}_{s'+}(\tau'),  \mathscr{S}_+(n,z) ]]-[Q^{1}_{s'+}(\tau') , [ Q^{2}_{s+}(\tau),  \mathscr{S}_+(n,z) ]]\cr
&=[  [Q^{1}_{s+}(\tau) ,Q^{2}_{s'+}(\tau') ],\mathscr{S}_+(n,z) ]-[  [Q^{1}_{s'+}(\tau') ,Q^{2}_{s+}(\tau) ],\mathscr{S}_+(n,z) ]\cr
&=\pi \frac{\kappa^4 }{4}(-)^{s+s'}
 D^{s+s'+1} \left[(s'+1) \tau' D_z\tau -(s+1) \tau D\tau'\right] (z)\d_{s+s'-1,n}\,,
\ee
from which
\be
[Q_{s}(\tau), Q_{s'}(\tau')]_+^1&=[Q^{1}_{s+}(\tau), Q^{2}_{s'+}(\tau')]+[Q^{2}_{s+}(\tau), Q^{1}_{s'+}(\tau')]\cr
&=
 \frac{\kappa^2}{{4}}\left[(s'+1) Q^{1}_{s+s'-1+}(\tau' D \tau) 
-(s+1) Q^{1}_{s+s'-1+}(\tau D \tau') 
\right]\,.
\ee

\subsection{Quadratic order global}\la{App:GRalg-qua-g}

In this section with give the proof of the relation \eqref{Q2Q2}. 
We start with the proof of \eqref{magic-g} for global charges. Using the  generalised Leibniz rule \eqref{magicYM}, we  find that 
\be
 \alpha D^{\alpha-1} (D\tau D^{s-\alpha} ) &=  \sum_{n=0}^\infty \frac{\alpha  (\alpha-1)_{n}}{ n! }   (D^{n+1} \tau ) D^{s-n-1}\cr
 &= \sum_{n=0}^\infty \frac{(\alpha)_{n+1}}{ n! }   (D^{n+1} \tau ) D^{s-n-1}\cr
&=  \sum_{n=0}^\infty  n \frac{(\alpha)_{n}}{ n! }   (D^{n} \tau ) D^{s-n}.
\ee
Summing these two contributions,  we find that we have the key identity 
\be
 D^{\alpha-1} \left[(s+1) \tau D +(s+1 -  \alpha) D\tau\right] D^{s-\alpha}
 = \sum_{n=0}^\infty  (s+1-n) \frac{(\alpha)_{n}}{ n! }   (D^{n} \tau ) D^{s-n}.
\ee
If we demand that 
\be
D^{s+2} \tau_s =0, 
\ee
we see that the sum can be restricted to the range $s \leq n$. In this case all the derivative operators appear with a positive power. There is no longer any non-locality. This means that the charge action \eqref{Q2Nn} can therefore be simply written as
\eqref{R1}.

We then use this result to compute
\be
&[Q^2_{s',+}(\tau') , [ Q^2_{s,+}(\tau),   \mathscr{S}_+(n,z) ]]_G
\cr
&=
-\f{\kappa^2}{{4}}(-)^{s} 
 D_z^{n+2} \left[(s+1) \tau D_z +(s-n-2) D\tau\right] D_z^{s-n-3}
[Q^2_{s',+}(\tau') , \mathscr{S}_+(n-s+1,z)]_G
\cr
&=\kappa^4 \frac{(-)^{s+s'} }{{16}}  
 D_z^{n+2} \left[(s+1) \tau D_z +(s-n-2) D\tau \right] 
  \left[(s'+1) \tau' D_z +(s+s'-n-3) D\tau'\right] \cr
&\times  D_z^{s+s'-n-4}  \mathscr{S}_+(n-s-s'+2,z)\,,
\ee
{where we used \eqref{Q2Nn2} twice.}
We now expand
\be
&\left[(s+1) \tau D +(s-n-2) (D\tau)\right] 
  \left[(s'+1) \tau' D +(s+s'-n-3) (D\tau')\right]
  -(s\leftrightarrow s', \tau \leftrightarrow \tau')\cr
&  =
 (s+1)  (s'+1)  \left[ \tau( D  \tau') D+  \tau  \tau' D^2\right] +(s+1)(s+s'-n-3)  \left[ \tau D^2\tau'+  \tau  (D\tau' ) D\right]\cr
&  +(s-n-2)(s'+1)(D\tau)  \tau' D 
+(s-n-2)(s+s'-n-3)(D\tau)  (D\tau')
-(s\leftrightarrow s', \tau \leftrightarrow \tau')
\cr
&=(1 + s) (s + s')\tau( D  \tau') D
 -(1 + s') (s + s')\tau' (D\tau)   D \cr
&+(s+s'-n-3) \left[ (s+1)  \tau D^2\tau'
-(s'+1) \tau' D^2 \tau  +(s - s') (D\tau)  (D\tau') \right]\cr
&= (s+s') [\tau,\tau'] D +(s+s' -  n-3) D[\tau,\tau'],\label{brabra}
\ee
where  we defined
\be
[\tau,\tau'] := (s+1) \tau D\tau' - (s'+1) \tau' D\tau\,.
\ee
Therefore, we have 
\be
&[Q^2_{s'+}(\tau') , [ Q^2_{s+}(\tau),   \mathscr{S}_+(n,z) ]]_G-[Q^2_{{s}+}({\tau}) , [ Q^2_{{s'}+}({\tau'}),   \mathscr{S}_+(n,z) ]]_G \cr
& =\kappa^4 \frac{(-)^{s+s'} }{{16}}   \bigg[
(s+s')  D^{n+3} \left[ [\tau,\tau']  D^{s+s'-n-4}  \mathscr{S}_+(n-s-s'+2,z)\right]  \cr
&  -(n+3)  D^{n+2} \left[ D[\tau,\tau']  D^{s+s'-n-4}  \mathscr{S}_+(n-s-s'+2,z)\right] 
\bigg]
\cr
& =\kappa^4  \frac{(-)^{s+s'} }{{16}}  
\bigg[
(s+s')   \sum_{k=0}^{n+3} 
 \left(\begin{matrix}
n+3\\
k
\end{matrix}\right)
 \left[D^k  [\tau,\tau']  D^{s+s'-1-k}  \mathscr{S}_+(n-s-s'+2,z)\right]  \cr
&  -(n+3)   \sum_{k=0}^{n+2} 
 \left(\begin{matrix}
n+2\\
k
\end{matrix}\right)
 \left[ D^{k+1}[\tau,\tau']  D^{s+s'-1-(k+1)}  \mathscr{S}_+(n-s-s'+2,z)\right] 
\bigg]
\cr
& =\kappa^4  \frac{(-)^{s+s'} }{16}  
   \sum_{k=0}^{n+3} 
 (s+s'-k) \left(\begin{matrix}
n+3\\
k
\end{matrix}\right)
D^k  [\tau,\tau']  D^{s+s'-1-k}  \mathscr{S}_+(n-s-s'+2,z) \,.\la{Q2Q2-app}
\ee

We conclude by verifying  the action
\be
&- \frac {\k}{{4}}\left[(s'+1)[ Q^{2}_{s+s'-1,+}(\tau' D \tau) , \mathscr{S}_+(n,z) ]-(s+1) [Q^{2}_{s+s'-1,+}(\tau D \tau') , \mathscr{S}_+(n,z) ]
\right]\cr
&=-\kappa^4 \frac {(-)^{s+s'}}{{16}}(s'+1)\sum_{k=0}^{s+s'-1} 
(s+s'-k)  \left(\begin{matrix}
n+3\\
k
\end{matrix}\right)
  D^{k} (\tau' D \tau)
  D^{s+s'-1-k} \mathscr{S}_+(n-s-s'+2,z)  \cr
  &-(s\leftrightarrow s', \tau \leftrightarrow \tau')\,,
\ee
which matches \eqref{Q2Q2-app} up to a sign, and thus \eqref{Q2Q2g} follows.

\subsection{Quadratic order remainders}\la{App:algebra-quadratic-22}

In this section we compute the two remainder nested commutators  \eqref{Rem}.
By means of \eqref{R2grav}, we compute first
\be\la{R2try2}
&[Q^{2}_{s'+}(\tau') , [ Q^{2}_{s+}(\tau),  \mathscr{S}_+(n,z) ]_{R}]
\cr
&=\frac{(-)^{s} \k}{{4}}    \sum_{k=s+1}^{n+3}
 (s-k+1 ) 
 \left(\begin{matrix}
n+3\\
k
\end{matrix}\right)
  D^{k} \tau(z)
  D^{s-k}[Q^2_{s',+}(\tau'), \mathscr{S}_+(n-s+1,z)]
  \cr
  &=-\frac{(-)^{s+s'} \kappa^4}{{16}}    \sum_{k=s+1}^{n+3}
   \sum_{\ell=0}^{s'}
   (s-k+1 ) 
 \left(\begin{matrix}
n+3\\
k
\end{matrix}\right)
 (s'-\ell+1 ) 
 \left(\begin{matrix}
n-s+4\\
\ell
\end{matrix}\right)
\cr
&\times
  D^{k} \tau(z)
  D^{s-k}
  \left[
  D^{\ell} \tau'(z)
  D^{s'-\ell}\mathscr{S}_+(n-s-s'+2,z)
  \right]
  \cr
&=-\frac{(-)^{s+s'} \kappa^4}{{16}}    \sum_{k=s+1}^\infty
   \sum_{\ell=0}^{s'}  \sum_{m=0}^{\infty}
   (s-k+1 ) 
 \left(\begin{matrix}
n+3\\
k
\end{matrix}\right)
 (s'-\ell+1 ) 
 \left(\begin{matrix}
n-s+4\\
\ell
\end{matrix}\right)
(-)^m \f{(k-s+m-1)!}{m!(k-s-1)!}
\cr
&\times
  D^{k} \tau(z)
D^{\ell+m} \tau'(z)
  D^{s+s'-\ell-k-m}\mathscr{S}_+(n-s-s'+2,z),
\ee
where we used the generalized Leibniz rule
\be
 &D^{s-k}
  \left[
  D^{\ell} \tau'(z)
  D^{s'-\ell}\mathscr{S}_+(n-s-s'+2,z)
  \right]
  \cr
  &=
   \sum_{m=0}^{\infty}  \f{(s-k)_m}{m!}D^{\ell+m} \tau'(z)
  D^{s'-\ell+s-k-m}\mathscr{S}_+(n-s-s'+2,z),
  \ee
  with
  \be
\f{(s-k)_m}{m!}
=\f{(-(k-s))_m}{m!}
=(-)^m \f{(k-s+m-1)_m}{m!}.
\ee

The other contribution that we want to cancel is
\be\la{R1R2}
&[Q^{2}_{s'+}(\tau') , [ Q^{2}_{s+}(\tau),  \mathscr{S}_+(n,z) ]_{G}]_{R}
\cr
&:=
-\frac{(-)^{s} \k}{{4}} 
 D^{n+2} \left[(s+1) \tau D +(s -  n-2) D\tau\right] D^{s-n-3}[Q^2_{s',+}(\tau'),\mathscr{S}_+(n-s+1,z)]_{R}
 \cr
 &=  -\frac{(-)^{s+s'} \kappa^4}{{16}} 
  \sum_{k=s'+1}^\infty 
 (s'-k+1 ) 
 \left(\begin{matrix}
n-s+4\\
k
\end{matrix}\right)
\cr
&\times
 D^{n+2} \left[(s+1) \tau D +(s -  n-2) D\tau\right] D^{s-n-3}[
   D^{k} \tau'(z)
  D^{s'-k}\mathscr{S}_+(n-s-s'+2,z)
 ]
 \cr
 &= -\frac{(-)^{s+s'} \kappa^4}{{16}} 
 \sum_{k=s'+1}^{\infty}
   (s'-k+1 ) 
 \left(\begin{matrix}
n-s+4\\
k
\end{matrix}\right)\cr
&\times
\bigg(
(s+1)  D^{n+2} \left[ \tau  D^{s-n-2}
 \left[ D^{k} \tau'(z)
  D^{s'-k}\mathscr{S}_+(n-s-s'+2,z)\right]\right]
  \cr
  &
 +(s-n-2) D^{n+2} \left[D\tau D^{s-n-3}
 \left[ D^{k} \tau'(z)
  D^{s'-k}\mathscr{S}_+(n-s-s'+2,z)\right]\right]
  \bigg)
  \cr
 &= -\frac{(-)^{s+s'} \kappa^4}{{16}} 
 \sum_{k=s'+1}^{n-s+4}\sum_{m=0}^{n+2}
   (s'-k+1 ) 
 \left(\begin{matrix}
n-s+4\\
k
\end{matrix}\right)
\left(\begin{matrix}
n+2\\
m
\end{matrix}\right)
\cr
&\times
\bigg(
(s+1) D^m \tau  D^{s-m}
 \left[ D^{k} \tau'(z)
  D^{s'-k}\mathscr{S}_+(n-s-s'+2,z)\right]
  \cr
  &
 +(s-n-2) D^{m+1}\tau D^{s-1-m}
 \left[ D^{k} \tau'(z)
  D^{s'-k}\mathscr{S}_+(n-s-s'+2,z)\right]\
    \bigg).
\ee

We conclude with the derivation of the  \eqref{Q1Q3S}. The action \eqref{Q3S} simplifies to
\be
[ Q^{3}_{2+}(\tau),  \mathscr{S}_+(n,z) ]
=-\frac{3 \k}{16\pi}  \sum_{k=0}^{n} 
 \left(\begin{matrix}
n+1\\
k
\end{matrix}\right)
 \tau_{2}(z)  \mathscr{S}_+(n-k, z) 
\mathscr{S}_+(k, z).
\ee
From this, we can then compute
\be
&[Q^{1}_{s'+}(\tau') , [ Q^{3}_{2+}(\tau),  \mathscr{S}_+(n,z) ]]
\cr
& =
- \frac{3 (-)^{s'}\k}{16\pi} 
 \sum_{k=0}^{\infty} 
 \left(\begin{matrix}
n+1\\
k
\end{matrix}\right)
 \tau_{2}(z) 
  \int_S  \rd {z'}^2 \sqrt{q}\, D_{z'}^{s'+2} \tau'(z') [ \mathscr{M}^\dagger_+(s',z') , \mathscr{S}_+(n-k, z) 
\mathscr{S}_+(k, z) ]
\cr
&=
 \frac{3 (-)^{s'}\kappa^4}{16} 
\left[
 \left(\begin{matrix}
n+1\\
n-s'
\end{matrix}\right)
 +
  \left(\begin{matrix}
n+1\\
s'
\end{matrix}\right)
\right]
\tau_{2}(z) D^{s'+2} \tau'(z) \mathscr{S}_+(n-s', z)
\cr
&=
 \frac{3 (-)^{s'}\kappa^4}{16} 
 \left(\begin{matrix}
n+2\\
s'+1
\end{matrix}\right)
\tau_{2}(z) D^{s'+2} \tau'_{s'}(z) \mathscr{S}_+(n-s', z).
\ee

\bibliographystyle{bib-style2.bst}
\bibliography{biblio-w.bib}

\end{document}